\documentclass[11pt,a4paper]{article}
\usepackage{jheppub,amsmath,  amssymb,slashed,url,bm,textgreek,upgreek}
\usepackage{graphicx}
\usepackage{epstopdf}
\def\t{{ \widetilde t}} 

\def\tt{{\mathfrak t}}

\def\s{{\mathfrak s}}

\def\p{{\eurm p}}

\def\be{\begin{equation}}
\def\ee{\end{equation}}

\def\h{\widehat}

\def\O{{\mathcal O}}

\def\Bbb{\mathbb}

\def\A{{\mathcal A}}
\def\d{{\mathrm d}}

\def\R{{\mathbb R}}
\def\C{{\mathbb C}}

\def\[{\bigl [}

\def\]{\bigr ]}

\def\cT{{\mathcal T}}

\def\tr{{\mathrm {tr}}}
\def\Z{{\mathbb Z}}

\def\t{\widetilde }
\def\h{\widehat}

\def\P{{\mathcal P}}

\def\H{{\mathcal H}}

\def\bar{\overline}

\font\teneurm=eurm10 \font\seveneurm=eurm7  \font\fiveeurm=eurm5
\newfam\eurmfam
\textfont\eurmfam=\teneurm \scriptfont\eurmfam=\seveneurm
\scriptscriptfont\eurmfam=\fiveeurm

\font\teneusm=eusm10 \font\seveneusm=eusm7 \font\fiveeusm=eusm5
\newfam\eusmfam
\textfont\eusmfam=\teneusm \scriptfont\eusmfam=\seveneusm
\scriptscriptfont\eusmfam=\fiveeusm

\font\tencmmib=cmmib10 \skewchar\tencmmib='177
\font\sevencmmib=cmmib7 \skewchar\sevencmmib='177
\font\fivecmmib=cmmib5 \skewchar\fivecmmib='177
\newfam\cmmibfam
\textfont\cmmibfam=\tencmmib \scriptfont\cmmibfam=\sevencmmib
\scriptscriptfont\cmmibfam=\fivecmmib

\def\sT{{\sf T}}
\def\cR{{\mathcal R}}
\def\sR{{\sf R}}
\def\sC{{\sf C}}

\def\CPT{{\sf CPT}}
\def\CRT{{\sf CRT}}
\def\sP{{\sf P}}
\def\la{{\langle}}
\def\ra{{\rangle}}
\def\veps{\varepsilon}
\def\i{{\mathrm i}}

\def\sE{{\sf E}}
\def\sF{{\sf F}}

\def\SO{{\rm{SO}}}
\def\O{{\rm O}}
\def\Pin{{\rm{Pin}}}
\def\Spin{{\rm{Spin}}}
\def\p{{\mathfrak p}}
\def\Hol{{\rm{Hol}}}

\title{Bras and Kets in Euclidean Path Integrals}

 \author{Edward Witten}
\affiliation{School of Natural Sciences, Institute for Advanced Study,\\ 1 Einstein Drive, Princeton, NJ 08540 USA}
\abstract{Here we discuss the relation between bras and kets in Euclidean path integrals.   In a theory lacking time-reversal or reflection symmetry,
bras and kets differ by an operator $\cT$ that complex conjugates the wavefunction and reverses the orientation of spacetime.   In the presence of time-reversal
and reflection symmetry, space is unoriented so $\cT$ cannot be defined, but the time-reversal symmetry $\sT$ is available instead.
An important special case of this concerns the relation between the hermitian inner product $\la ~,~\ra$ -- linear in one variable, antilinear in the other
-- that is used to define quantum mechanical probabilities, and the symmetric bilinear operator $(~,~)$ that comes most naturally from the Euclidean path integral.
They differ by the action of $\cT$ or $\sT$ on one side.
}

\begin{document}\maketitle

\section{Introduction}\label{intro}

In a scattering amplitude, initial states are mapped to final states.  Somewhat similarly, an operator matrix element 
$\la \Psi|\O|\chi\ra$ has an  initial state $\chi$ and a final state $\Psi$. The matrix element is linear in $\chi$ and antilinear in $\Psi$, because 
$\la~|~\ra$ is a hermitian form, not a bilinear one.

\begin{figure}
 \begin{center}
   \includegraphics[width=2.8in]{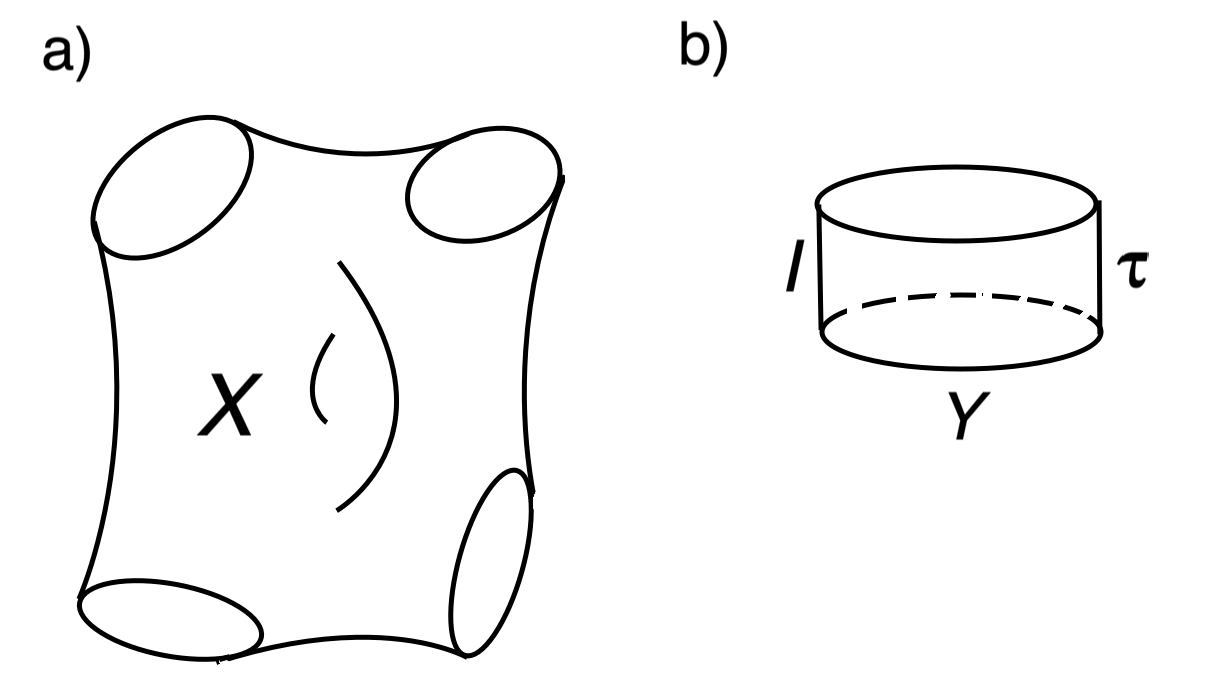}
 \end{center}
\caption{\footnotesize    (a) A Euclidean $D$-manifold $X$ with $n$ boundary components, depicted here with $n=4$. The amplitude has bose/fermi symmetry
in the external closed universe states.  (b) The special case that $n=2$
and $X=Y\times I $, where $Y$ is a $D-1$-manifold and $I$ is an interval of width $\tau$. In quantum field theory without gravity, the bilinear pairing
$(~,~)$ is defined by this path integral in the limit $\tau\to 0$.  In topological field theory, there is no dependence on $\tau$.
 In a gravitational theory, one has to project such a path integral  onto states that obey the gravitational constraints.  \label{one}}
\end{figure}

In this article, we will consider  amplitudes coupling external states in Euclidean path integrals.   We consider both closed and open universes,
but conceptually the case of closed universes is particularly interesting.
In fig. \ref{one}(a), we schematically depict a Euclidean path integral amplitude on a $D$-manfold $X$ with $n$ closed universe boundaries on which external
states can be defined.   Such  amplitudes makes sense in any quantum field theory, but are of particular interest in topological field theories and also in the context of quantum gravity.
An amplitude of this type, defined in the most natural way, is linear in each of the external states, and treats them symmetrically (so the amplitude is
invariant  under permutations of the external closed universes to the extent that such permutations extend to symmetries of the bulk manifold $X$; in gravity, such
symmetry is achieved after summing over all choices of $X$).
In such a picture there is  no visible distinction between quantum mechanical bras and kets.  In particular, if we set $n=2$ as in fig. \ref{one}(b), the Euclidean amplitude is bilinear in the two states, rather than being a hermitian pairing
(linear in one state and antilinear in the other) like a quantum mechanical matrix element.

In the present article, we will analyze how to incorporate bras and kets in Euclidean path integrals. In particular, we will see what one should do in fig. \ref{one}(b) if one wishes
to compute a hermitian pairing between the two states rather than a bilinear pairing.

As one might expect, the answer involves the behavior of the quantum wavefunction under complex conjugation.   In detail, there are two somewhat distinct cases depending on
whether the theory is assumed to possess a symmetry of spatial reflection or time-reversal.   
The Standard Model of particle physics is an example of a theory with no such symmetry.   In the absence of reflection symmetry,   
to define a quantum state on a spatial manifold $Y$, in additional to specifying a functional of the fields on  $Y$, one has to pick an orientation of $Y$.  
In a closed universe, this leads to  a sort of doubling of the Hilbert space that is not always taken into account (in an open universe, this doubling can be avoided
by picking once and for all an orientation at spatial infinity).    Roughly speaking, in the case of a theory with no time-reversal or reflection 
symmetry, a bra differs from a ket by complex conjugating the wave function
and reversing the orientation of space.  Theories with a  reflection and time-reversal symmetry are more subtle.
Roughly speaking, in such a theory there is no doubling associated with a choice of orientation -- indeed, such theories can be formulated on unorientable manifolds -- and a bra 
is obtained from a ket by acting with time-reversal.

In section \ref{sectwo}, we discuss the relation between bras and kets in the relatively straightforward case of a  theory that has  no reflection symmetry.   After presenting in section \ref{background} some general background on reflection  and time-reversal symmetries,
we describe in section \ref{secfour}  the relation between bras and kets in  theories with such symmetries. 
Actually, the discussion of time-reversal and reflection symmetries is more straightforward in the absence of fermions, so in section \ref{secfour}
we consider bosonic theories only.   Fermions are included in section \ref{incfermions}.  The antilinear nature of $\CPT$ or $\CRT$ symmetry is discussed in an appendix.

$\CRT$ symmetry in gravity has been recently discussed from another point of view in \cite{Harlow}.  A number of topics considered in the present article, such as reflection 
positivity in quantization on possibly unorientable manifolds, possibly with fermions,  have been analyzed previously with much more mathematical
 detail in \cite{FH}.   A classic reference on $\CPT$ symmetry and related matters from an axiomatic point of view is \cite{StW}.

\section{Theories With No Reflection Symmetry}\label{sectwo}

\subsection{An Example}

A useful illustrative example of a quantum field theory with no reflection or time-reversal symmetry is a four-dimensional gauge theory with a theta-angle.   The action in Lorentz signature is
\be\label{lorac}I_L =-\frac{1}{4e^2}\int \d^4x \sqrt g\, \tr\,F_{\mu\nu} F^{\mu\nu} +\frac{\theta}{16\pi^2}\int\d^4x \sqrt g \epsilon^{\mu\nu\alpha\beta}\tr\,F_{\mu\nu}F_{\alpha\beta},\ee
where $F$ is the gauge field and gauge field strength, $e$ and $\theta$ are the gauge coupling constant and theta-angle, $g$ is the spacetime  metric tensor with signature $-+++$, $\epsilon_{\mu\nu\alpha\beta}$
is the Levi-Civita antisymmetric tensor, and 
(for the case of a gauge group ${\rm{SU}}(N)$ or ${\rm U}(N)$) $\tr$ is the trace in the fundamental representation.   For an illustrative example of a theory of gravity with no reflection or time-reversal
symmetry, we can simply take Einstein gravity coupled to a gauge theory with a theta-angle.   A canonical example of a topological field theory with no reflection or time-reversal symmetry
is three-dimensional Chern-Simons theory with  a simple gauge group.

\begin{figure}
 \begin{center}
   \includegraphics[width=1.3in]{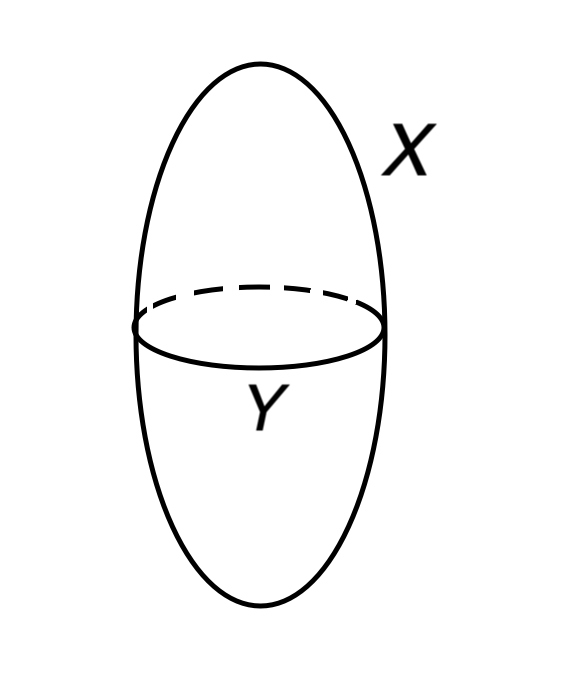}
 \end{center}
\caption{\footnotesize   $X$ is a reflection-symmetric manifold that can be divided in two parts by cutting on its plane of symmetry $Y$.   For any
fixed values of the fields on $Y$, the path integral on the part of $X$ above the cut is the complex conjugate of the path integral on the part below the cut, because
the reflection symmetry of $X$ reverses the orientation of $X$ and complex conjugates the action and the path integral. 
Hence for any values of the fields on $Y$, the product of the path integrals above and below $Y$ is positive.   Upon integrating over fields on $Y$, this  implies that
the path integral on $X$ is positive.   This property, known as reflection positivity, implies that the norm of the state on $Y$ that is prepared by the path integral
on the lower portion of $X$ is positive.   This is the basic argument proving positivity of Hilbert space norms in the context of path integrals.  \label{refpos}}
\end{figure}

In Lorentz signature, the action of any unitary theory is always real; in particular, in (\ref{lorac}) the parameters $e$ and $\theta$ are real.   
In Euclidean signature, the action of a theory that lacks reflection symmetry is  not real.  Indeed, when we transform from Lorentz to Euclidean signature, the metric tensor remains real, but the Levi-Civita tensor
$\epsilon_{\mu\nu\alpha\beta}$ acquires a factor of $\i$.   Thus the Euclidean action is
\be\label{zorac} I_E=\frac{1}{4e^2}\int \d^4x \sqrt g \tr\,F_{\mu\nu} F^{\mu\nu} -\frac{\i\theta}{16\pi^2}\int\d^4x \sqrt g \epsilon^{\mu\nu\alpha\beta}\tr\,F_{\mu\nu}F_{\alpha\beta}.\ee
We can make a simple general statement that would apply for any Lorentz-invariant action.
The parity-violating terms are proportional to an odd power of $\epsilon_{\mu\nu\alpha\beta}$, so they are imaginary in Euclidean signature, while the parity-conserving terms
are proportional to an even power of $\epsilon_{\mu\nu\alpha\beta}$, and are real.
Accordingly  under a reversal
of orientation (which changes the sign of $ \epsilon_{\mu\nu\alpha\beta}$), the Euclidean action is complex-conjugated.  
In a theory with fermions, the analysis is more complicated but the conclusion is the same: the Euclidean action in a purely bosonic theory,
and  the effective action after integrating out fermions in general,  is complex-conjugated under
reversal of orientation.\footnote{In general, in Euclidean signature, fermions do not satisfy any reality condition and even in the absence of parity violation, one cannot claim that the Euclidean
action for fermions is real.  A counterexample is the theory of a single Majorana fermion in four dimensions; the theory is parity-conserving but because a Majorana fermion Wick rotates
to a field that is in a pseudoreal (rather than real) representation of the rotation group, there is no sense in which
the Euclidean action is real. In this particular example, because the representation is
pseudoreal, the fermion path integral is actually real
and does not depend on a choice of orientation.
 For a more subtle example,
 consider in two dimensions a one-component chiral fermion field $\lambda$, which rotates to a field that in Euclidean signature is in a complex representation.
 In this case, the Euclidean path integral is actually complex (and in fact, it suffers from a gravitational anomaly; to define it on a curved manifold, the theory of the field $\lambda$
 must be embedded in a larger anomaly-free theory).   Reversal of orientation reverses the chirality of $\lambda$ and complex conjugates the Euclidean path integral.}
   The fact that the path integral is complex-conjugated under reversal of orientation is essential in the
proof of reflection positivity and thus unitarity; see fig. \ref{refpos} for an explanation of this point.   For a thorough analysis of reflection positivity in topological field theory (where arbitrary
topologies have to be considered), in theories possibly with fermions and/or  parity violation, 
see \cite{FH}.

A parity-violating theory cannot be defined on an unorientable manifold.     It can only be defined on a manifold $X$
that is not just orientable but oriented, that is, a manifold on which an orientation has been chosen.  We write $\veps$ for the orientation of $X$.
   Concretely one can think of $\veps$ as a choice of sign of the tensor
$\epsilon_{\mu\nu\alpha\beta}$.   Now write  $Z_X(g,\veps)$ for the partition function of a given theory on a manifold $X$ with a chosen metric $g$ and orientation $\veps$.
Denote the orientation opposite to $\veps$ as $-\veps$.   Reversing the orientation complex conjugates the action, so it complex conjugates the argument $e^{-I_E}$ of the
Euclidean path integral and therefore complex conjugates the partition function.   Hence the partition function has a simple dependence on the orientation:
\be\label{milko} Z_X(g,\veps)=\overline {Z_X}(g,-\veps). \ee
Of course, in a topological field theory (in which there is no dependence of physical observables on $g$) or a gravitational theory (in which we eliminate the dependence on $g$ by
integrating over $g$), the partition function depends on $\veps$ only, and not on $g$.

\subsection{Orientations and Hilbert Spaces}\label{orhilb}

Now we will study the physical states obtained in quantizing a $D$-dimensional theory that has no reflection symmetry on a spatial manifold $Y$ of dimension $D-1$.
Naturally, in the absence of a reflection symmetry, the Hilbert space of physical states will depend on the orientation of $Y$. 
If $X$ is a Euclidean $D$-manifold with boundary $Y$, then a path integral on $X$, as a function of fields $\phi$ on $Y$, computes a physical state $\Psi(\phi)$.
In Euclidean signature,  a reversal of the orientation of $X$ will complex conjugate the   action $I_E$ on $X$ and the path integral on $X$. So the state $\Psi(\phi)$ on $Y$
is also complex conjugated.  This example indicates that  there is a natural operation on physical states that reverses the orientation of $Y$ and complex
conjugates the wavefunction.

It is instructive see this explicitly in the context of
the gauge theory (\ref{lorac}) with a theta-angle.   To quantize the theory, we work on $\R\times Y$,  with a Lorentz signature metric
$-\d t^2+\sum_{i,j=1}^3 h_{ij}\d x^i \d x^j$, where $t=x^0$ parametrizes $\R$, and $x^i$ are local coordinates on $Y$.  Given a Levi-Civita tensor $\epsilon_{\mu\nu\alpha\beta}$ of $X=\R\times Y$,
we define a Levi-Civita tensor $\t\epsilon_{ijk}=\epsilon_{0ijk}$ on $Y$.  

Just as the path integral on $X$ depends on a choice of orientation of $X$, the Hamiltonian formalism on $Y$ depends on a choice of orientation of $Y$.  To see this explicitly, we first
compute the canonical momentum:
\be\label{conjmom}\Pi_i =\frac{1}{ e^2} F_{0i}+\frac{\theta}{8\pi^2} \t\epsilon_{ijk} F^{jk}\ee
and the Hamiltonian
\begin{align}\label{hamton} H&=\int_Y\d^3x \sqrt h\,\tr \left(\frac{1}{2e^2} \tr\,F_{0i}(x)^2+\frac{1}{4e^2} \tr\,F_{ij}F^{ij}\right)\cr &
=\int_Y\d^3x\sqrt h\,\tr\left(\frac{e^2}{2} \left(\Pi_i(x)-\frac{\theta}{8\pi^2} \t\epsilon_{ijk}F^{jk}(x)\right)^2+\frac{1}{4e^2} \tr\,F_{ij}F^{ij}\right) . \end{align}
In the gauge $A_0=0$, a physical state is a gauge-invariant function\footnote{Formally, the space of gauge-invariant square-integrable functions $\Psi(A_i(x))$ does not depend on 
the orientation of space, so in this theory the Hilbert space formally does not depend on orientation, though the Hamiltonian and other operators do, as we are about to see.
After coupling to gravity, the Hilbert space of this theory does depend on orientation, because it is defined with the help of gravitational constraint equations that depend on the
energy and momentum densities.  Reversing the orientation complex conjugates the energy and momentum operators, so it complex conjugates the Hilbert space.
   Examples of theories without gravity whose Hilbert spaces depend on orientation are the Standard Model (whose Hilbert space
depends on the orientation of space because of fermion chirality), gauge theories in three dimensions with Chern-Simons couplings, and sigma-models with Wess-Zumino terms. In
all cases, orientation-reversal complex conjugates the Hilbert space, meaning that it exchanges bras and kets.}
 $\Psi(A_i(x))$.  Upon canonical quantization, $\Pi_i(x)$ is replaced by $-\i\frac{\delta}{\delta A_i(x)}$, so 
the Hamiltonian acts by
\be\label{amton}H=\int_Y\d^3x \sqrt h\,\tr \left(\frac{e^2}{2}\left(-\i\frac{\delta}{\delta A_i(x) }-\frac{\theta}{8\pi^2} \t\epsilon_{ijk}F^{jk}(x)\right)^2 +\frac{1}{4e^2} \tr\,F_{ij}F^{ij}\right) .\ee

The Hamiltonian depends on 
the Levi-Civita tensor $\t\epsilon$, or equivalently, it depends on the orientation $\veps_Y$ of $Y$.   Thus in quantizing the theory, we need to specify
the orientation $\veps_Y$ as well as the metric $g$ of $Y$.    We will leave implicit  the metric
dependence (which is absent in a topological field theory, while in  a theory of gravity the metric is one of the fields the wavefunction depends on)  and denote the Hilbert space as $\H_{Y,\veps_Y}$.  

We see immediately that $H$ is hermitian but not real, and if we reverse the sign of $\t\epsilon$, then $H$ is complex conjugated.
   The Hamiltonian operator acting on $\H_{Y,\veps_Y}$ is thus the complex conjugate of the Hamiltonian operator acting on $\H_{Y,-\veps_Y}$.   
Accordingly, there  is a natural antilinear 
map $\cT:\H_{Y,\veps_Y}\to \H_{Y,-\veps_Y}$ between the two Hilbert spaces by complex conjugation of the wavefunction:
\be\label{pellme}\cT\Psi(A)=\bar\Psi(A). \ee
This map conjugates  the Hamiltonian acting on one Hilbert space to the Hamiltonian acting on the other.
The operator $\cT$ is antiunitary and satisfies
\be\label{ellme}\cT^2=1.  \ee
We have used the example of gauge theory with a theta-angle to illustrate the fact that orientation reversal should be accompanied by complex conjugation of the wavefunction.
However, the result is quite general.   This is perhaps obvious in Euclidean signature: for $\beta>0$, since the action is complex-conjugated under reversal of orientation, the
operator $e^{-\beta H}$, whose matrix elements are naturally computed by a Euclidean path integral, is complex-conjugated in a reversal of orientation, and therefore
the same is true for $H$.

Since complex conjugating  a hermitian operator such as the Hamiltonian
 is equivalent to taking its transpose, instead of saying that the Hamiltonian acting on $\H_{Y,\veps_Y}$ is the complex
conjugate of the Hamiltonian acting on $\H_{Y,-\veps_Y}$, an equivalent statement is that these operators are transposes of each other.  So if 
$H_{\veps_Y}$ is the Hamiltonian acting on $\H_{\veps_Y}$, and $H^{\rm tr}_{\veps_Y}$ is its transpose, then
\be\label{zellme}H^{\rm tr}_{\veps_Y}=H_{-\veps_Y}.\ee

For applications to topological field theory or gravity, it is important  that generically the orientable $D$-manifold $X$ or  $D-1$-manifold $Y$ is not equivalent to the same manifold
with the opposite orientation. Indeed, a generic orientable manifold (in dimension $\geq 3$) does not have any orientation-reversing diffeomorphism.   For example, in four dimensions, a manifold such as ${\mathbb {CP}}^2$
with a nonzero signature has no orientation-reversing diffeomorphism; similarly, in three dimensions the Chern-Simons invariant of an abelian flat connection can be used to prove
that a generic lens space $S^3/{\mathbb Z}_n$ does not have an orientation-reversing diffeomorphism.   We will return momentarily to the case that $Y$ does have an orientation-reversing
diffeomorphism.

The operation that we have called $\cT$, being antilinear,
 is somewhat reminiscent of time-reversal symmetry, but it is not what is usually meant by time-reversal symmetry; indeed, we are discussing a theory
with no time-reversal symmetry.  In fact, $\cT$ is not usually regarded as a symmetry.    For example, the Standard Model is usually quantized in Minkowski space with a particular choice
of spacetime orientation, defined perhaps by a ``right hand rule.''   Acting with $\cT$ would exchange this quantization with an equivalent quantization of the Standard Model based on a left
hand rule.   This is not usually considered useful; it tells us nothing about what happens when the Standard Model  is defined with the usual convention.

Matters are more subtle in the case of gravity.
In that context, the two cases of an open universe  and a closed universe are rather different.   
In semiclassical gravity, in
 an open universe, can pick an orientation once and for all at spatial infinity and that will always determine an orientation in the interior of the spacetime
 (or at least, the component of the spacetime that is connected to infinity).    There is no need to consider any other orientation and so
 $\cT$ plays no more  role than it does for the Standard Model in Minkowski space.

\begin{figure}
 \begin{center}
   \includegraphics[width=2.1in]{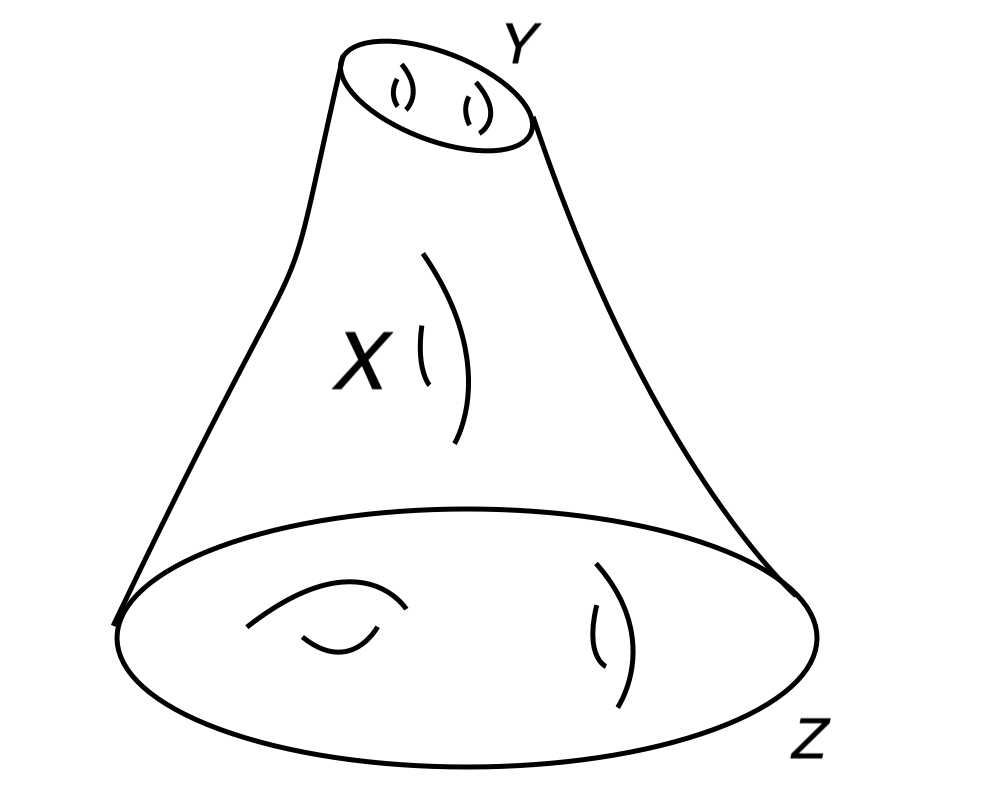}
 \end{center}
\caption{\footnotesize A state on a $D-1$-manifold $Y$ can be determined by asymptotic data at $Z$.   A familiar version of this occurs with negative cosmological  constant:
$X$ can have Euclidean signature and $Z$ is the conformal boundary of $X$.   With zero or positive cosmological constant, $X$ could be an expanding or contracting
FRW spacetime in Lorentz signature with $Z$ at the far future or past.   In all cases, with 
generic asymptotic data on $Z$, the induced state on $Y$ is not $\cT$-invariant.  \label{eucnot}}
\end{figure}

More interesting is the case of a closed universe.   In semiclassical quantization on closed universes, there is no natural 
way to avoid considering all orientable manifolds  $Y$ with all possible orientations on $Y$.
So, semiclassically, the Hilbert space is a direct sum 
\be\label{zelbo}  \H=\bigoplus_{Y,\veps_Y} \H_{Y,\veps_Y}.\ee
(For brevity, in writing this formula, we ignore the fact that some oriented  $d$-manifolds $Y$ do have orientation-reversing diffeomorphisms; for such manifolds,
we need not consider both choices of $\veps_Y$.)   
We see that the Hilbert space $\H$ of  closed universes has an antilinear symmetry $\cT$ that exchanges $\H_{Y,\veps_Y}$ with $\H_{Y,-\veps_Y}$. 
Thus, $\H$ has a real structure; a state in $\H$ is ``real'' if the part of the wavefunction in $\H_{Y,\veps_Y}$ is the complex conjugate of the part in $\H_{Y,-\veps_Y}$.
Should we treat $\cT$ as a gauge symmetry, that is, are states that are invariant under $\cT$ the only ones that we should consider to be physically sensible?
 It appears that the answer is ``no'' for the following reasons: (1) the gravitational path
integral does not have a mechanism involving a sum over ``twists'' of some sort that would project onto $\cT$-invariant states; (2) standard constructions via Euclidean 
path integrals construct states that are in general not $\cT$-invariant (fig. \ref{eucnot}) and there does not seem to be a good reason to reject such states.    However,
for   possible arguments for a  ``yes'' answer in holographic theories, see \cite{Harlow}.

The four-dimensional gauge theory that we have considered as an example, 
even though it does not have a parity or time-reversal symmetry, does, if formulated in Minkowski space, have a combined symmetry usually called $\sf{CPT}$
(we will introduce an alternative terminology in section \ref{background}).   In quantization on $\R^3$, $\sf{CPT}$ acts by  parity, $(x^1,x^2,x^3)\to (-x^1,-x^2,-x^3)$ combined with the
antiunitary operation of time-reversal.   One may ask what is an analog of $\sf{CPT}$ for quantum field theory (or topological field theory, or quantum gravity) on a curved manifold
$Y$.   The closest analog arises if $Y$ does have an orientation-reversing symmetry $\cR:Y\to Y$.  Of course, $\cR$ by itself is not a symmetry of a theory that lacks reflection symmetry,
precisely because it reverses the orientation.   But we can combine it with $\cT$, which also reverses the orientation, to get an antilinear but orientation-preserving operation
$\cR\cT$, which is a symmetry assuming the metric of $Y$ is $\cR$-invariant.    This operation, of course, is also a symmetry in a topological field theory or a theory of gravity.

    \begin{figure}
 \begin{center}
   \includegraphics[width=2.1in]{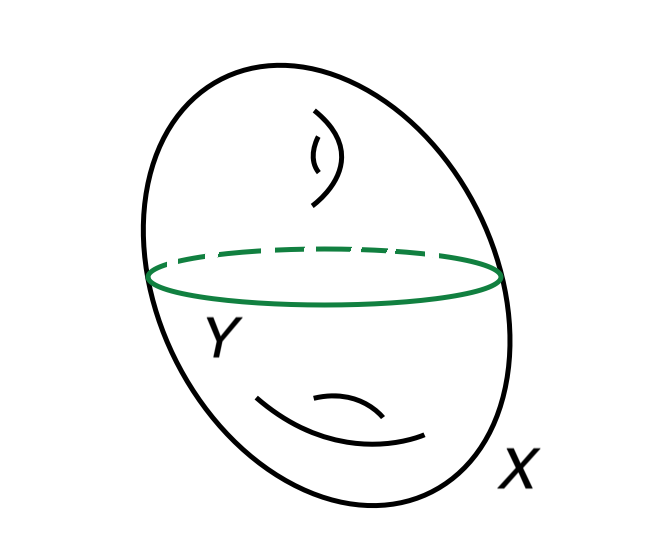}
 \end{center}
\caption{\footnotesize  A path integral on a $D$-manifold $X$ can be ``cut'' on any 
embedded $D-1$-manifold $Y$.  In a reflection-invariant theory of gravity, if $Y$ is compact and has an 
orientation-reversing symmetry $\rho$, one  can glue the two sides of the cut back together  after acting with $\rho$ on one side.   A sum over such ``twists'' will
project onto $\rho$-invariant states.  Hence only $\rho$-invariant states can propagate.   There is no close analog of this for $\cT$ or for the time-reversal symmetries considered
later.  \label{bucnot}}
\end{figure}

For the same reasons as in the discussion of $\cT$, it does not seem natural in the context of gravity to view $\cR\cT$ as a gauge symmetry.  Indeed, the Euclidean
path integral of fig. \ref{eucnot} can be used to prepare states that are not $\cR\cT$-invariant.

 In general, if $\cR$ and $\cR'$ are two different orientation-reversing symmetries
 of $Y$,   then $\cR\cT$ and $\cR'\cT$ are different as operators on $\H_{Y,\veps}$.  In topological field theory, $\cR\cT$ and $\cR'\cT$ are equivalent if $\cR $ and $\cR'$ are isotopic.
 In gravity in a closed universe, one expects that $\cR\cT$ does not depend on the specific choice of $\cR$.    The reason is that if $\cR$ and $\cR'$ are orientation-reversing,
 then $\cR'\cR^{-1}$ is orientation-preserving. In gravity in a closed universe,
 one would expect $\cR'\cR^{-1}$ to behave as a gauge symmetry, leaving physical states invariant (see fig. \ref{bucnot}).
In that case $\cR\cT$ and $\cR' \cT=(\cR'\cR^{-1})\cR\cT$ are equivalent.   

For closed universes, $\cT$  is the only universal antilinear symmetry that reverses the orientation of space
(albeit in a trivial way, by exchanging any given spatial manifold $Y$ with another oppositely oriented copy of itself).    So a case can be made for referring to $\cT$
as $\CRT$ (the combination of time-reversal and a spatial reflection with charge conjugation; see section \ref{background}).   We have not adopted this terminology here
as  in Minkowski space $\cT$ does not coincide with what is usually  called $\CRT$.   It is true, though, that $\cT$ is the only  antilinear operation that is available in quantization
on a generic spatial manifold.

\subsection{Bras and Kets and the Hermitian Inner Product}\label{hermin}

We are now prepared to answer the questions  raised in the introduction. In a theory that lacks time-reversal or reflection symmetry, 
what is the relation between bras and kets in a Euclidean path integral such as that of fig. \ref{one}(a)?
And in particular
what is the relationship between
the usual positive-definite hermitian inner product $\la~,~\ra$ on physical states, and the bilinear, permutation symmetric inner product $(~,~)$ that  comes most naturally from the
path integral of fig. \ref{one}(b)?   

For this purpose, we first reconsider the claim that the general amplitude of fig. \ref{one}(a), involving a path integral on a bulk
manifold $X$ with boundary components $Y_r$, $r=1,\cdots,n$  has permutation  symmetry among the external states.   The precise meaning of this statement depends on the type of theory
considered.  In a general quantum field theory,  the amplitude 
  is invariant under any permutations of the $Y_r$ that extend to  orientation-preserving isometries of $X$.   In topological field theory (where the metric of $X$ is irrelevant but the topology of $X$ is fixed), the  amplitude is invariant under any permutations of the $Y_r$
 that extend to orientation-preserving diffeomorphisms of $X$.  Finally, in quantum gravity,  the amplitude after summing over all choices of $X$  is potentially invariant
 under all permutations of the $Y_r$.

Let us consider  the amplitude of  fig. \ref{one}(a), taking orientations into account.   Once an orientation is picked on the
bulk manifold $X$, it induces orientations on all of the boundary components $Y_r$.     Concretely, along each $Y_r$, pick a Euclidean ``time'' coordinate $t_r$ that  is constant along $Y_r$ and increases toward the
interior of  $X$,
and pick local coordinates $x^1,x^2,\cdots x^d$ along $Y_r$.   Then if $X$ has been oriented with a choice of  Levi-Civita tensor that along $Y_r$ reduces to  $\d t_r\,\d x^1\cdots \d x^d$,
we orient $Y_r$ with the Levi-Civita form $\d x^1\d x^2 \cdots \d x^d$.   (Or we could reverse the sign and choose the orientation along $Y_r$ that corresponds to
$- \d x^1\d x^2\cdots \d x^d$.   The important thing is to use the same convention for all $X$ and  for all components $Y_r$.)   Permutations of the boundary components
automatically permute the induced orientations.

A special case that is frequently encountered in contemporary studies of semiclassical gravity (for example in applications of the replica trick to compute entropies) is that the boundary of $X$ consists of many copies of the same manifold $Y$.
 As boundary components of $X$,
the various copies of $Y$ all acquire orientations, but in general,
these  induced orientations will not agree.  Suppose that $n_1$ boundaries have one induced orientation and the other $n_2=n-n_1$ boundaries
have the opposite induced orientation.  This means that  external states  valued in $\H_{Y,\veps}$ (for some $\veps$) can be inserted on $n_1$ boundaries, and 
states valued in $\H_{Y,-\veps}$ can be inserted on the other $n_2$  boundaries.

Beyond this point, what one would say depends on the application one has in mind.
A point of view that is natural in gravity  is to consider the  amplitude of fig. \ref{one}(a) as a multilinear function on $n$ states valued in the direct sum
$\H_{Y,\veps}\oplus \H_{Y,-\veps}$.
This viewpoint is natural in the context of gravity, since $\H_{Y,\veps}$ and $\H_{Y,-\veps}$ both contribute to the closed universe Hilbert space (\ref{zelbo}).
We simply then insert on each boundary the appropriate part of the state, whether valued in $\H_{Y,\veps}$ or $\H_{Y,-\veps}$, depending on the induced orientation
of that boundary.  
After summing over all $X$ (and all orientations on each $X$), the amplitude understood this way has complete permutation symmetry in the external states.

Alternatively, what can we do if we want to consider all the external states to be valued in $\H_{Y,\veps}$, with the same orientation?
Such states can be inserted on the $n_1$  boundaries on which the induced orientation is $\veps$.   To insert $\H_{Y,\veps}$-valued states on the $n_2$ boundaries
with induced orientation $-\veps$,   we can first act with $\cT$ to map these states to $\H_{Y,-\veps}$, after which they can be inserted on the boundaries with that orientation.
The result is to get an amplitude that is linear in $n_1$ copies of $\H_{Y,\veps}$ and antilinear in $n_2$ additional 
copies of $\H_{Y,\veps}$.   In other words, this is an amplitude in which the external states consist of $n_2$ bras and $n_1$ kets.
Amplitudes of this kind (usually with $n_1=n_2$) are often considered
in calculations involving the replica trick (calculations that involve considering many copies of a quantum state).  

If we assume that in gravity the state is supposed to be $\cT$-invariant (contrary to what was argued previously in relation to fig. \ref{eucnot}), then applying $\cT$ to the part of the
state valued in $\H_{Y,\veps}$ just gives the part of the state valued in $\H_{Y,-\veps}$.   Once again, from this viewpoint, the sign of the induced orientation on the boundary distinguishes
bras and kets.

A simple and important example is the special case $X=I\times Y$ of fig. \ref{one}(b), where $Y$ is a general orientable $d$-manifold and $I$ a closed interval  that we parametrize by the Euclidean 
``time'' $t$, with $0\leq t\leq\tau$. We will take the limit $\tau\to 0$ to define inner products, but it is clearest to start with $\tau>0$.
In this example, no matter how we orient $X$, the induced orientations on the two boundaries are opposite.   Let $x^1,\cdots, x^d$ be local
coordinates on $Y$, and suppose, for example, that $X$ is oriented by the Levi-Civita form $\d t\, \d x^1 \cdots \d x^d$.   Then  $t$ increases in 
going away from one component but decreases away from the other, so the induced orientations on the two boundaries are opposite.   
Thus we can interpret this diagram as defining, in the limit $\tau\to 0$,  a bilinear form $(~,~):\H_{Y,-\veps}\otimes \H_{Y,\veps}\to \C$.   Alternatively, we can apply 
$\cT$ to the first state and get a hermitian pairing $\la~,~\ra:\H_{Y,\veps}\otimes \H_{Y,\veps}\to \C$. Clearly the relation between the two pairings is
\be\label{lombo}\la\Psi,\chi\ra=(\cT\Psi,\chi). \ee    Or we can apply $\cT$ to the second state
and get the complex conjugate hermitian pairing on the complex conjugate vector space $\H_{Y,-\veps}$.

   \begin{figure}
 \begin{center}
   \includegraphics[width=2.1in]{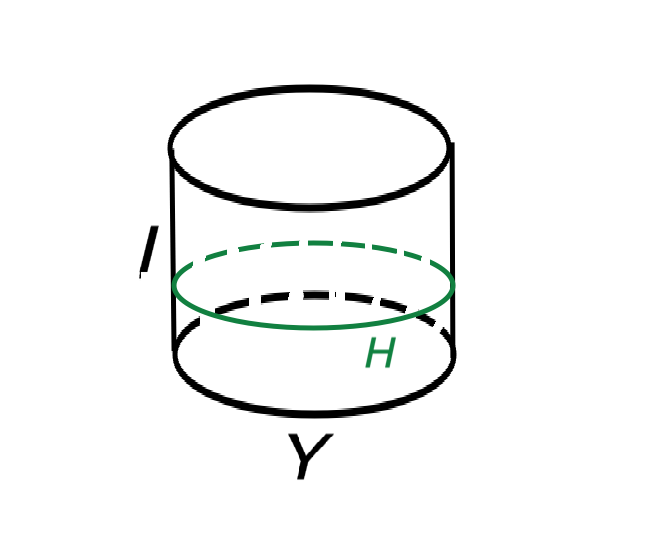}
 \end{center}
\caption{\footnotesize A path integral on $Y\times I$, where $I$ is an interval.  The Hamiltonian $H$ has been inserted on the submanifold   $Y\times p$, with $p$ a point in $I$.
As $H$ is conserved, it can be freely moved to the top or bottom of the cylinder, implying that $(H\Psi,\chi)=(\Psi,H\chi)$.   Thus $H$ is symmetric with respect to
$(~,~)$, as well as hermitian with respect to $\la~,~\ra$.  In a time-reversal invariant theory with time-reversal operator $\sT$ that satisfies $\sT^2=1$,
a similar argument with $\sT$ inserted
on $Y\times p$, taking into account that $\sT$ is antilinear,  proves that $(\sT\Psi,\chi)=\overline{(\Psi,\sT\chi)}$.   \label{deduc}}
\end{figure}

As a reality check, let us consider the
example of the gauge theory with a theta-angle.  In this example, $\Psi$ and $\chi$ are both functions of the gauge field $A$ on $Y$.  The bilinear pairing $(\Psi,\chi)$ that comes from the $\tau\to 0$ limit of the cylinder amplitude of fig. \ref{one}(b)  is formally  an integral over the gauge field $A$ on $Y$:
\be\label{dedo}(\Psi,\chi)=\frac{1}{{\rm vol}(\mathcal G)} \int DA \,\Psi(A)\chi(A),\ee
(here formally ${\rm vol}(\mathcal G)$ is the volume of the group of gauge transformations on $Y$) and the corresponding hermitian pairing is
\be\label{wedo}\la\Psi,\chi\ra =\frac{1}{{\rm vol}(\mathcal G)} \int DA\,\bar\Psi(A)\chi(A). \ee
However, in (\ref{dedo}), $\Psi(A)$ and $\chi(A)$ are vectors in Hilbert spaces $\H_{Y,-\veps_Y}$ and 
$\H_{Y,\veps_Y}$ defined with opposite orientations, while in (\ref{wedo}), they take values in the same Hilbert space.

We can see the necessity of this distinction between the Hilbert spaces as follows.
  Let $H$ be the Hamiltonian operator of eqn. (\ref{amton}).   Of course, it is supposed to be
hermitian with respect to $\la~,~\ra$:
\be\label{limbo}\la \Psi,H\chi\ra=\la H\Psi,\chi\ra. \ee
There is no problem with this; the expression for $H$ in eqn. (\ref{amton}) is indeed manifestly hermitian.    On the other hand,  one can deduce from the path integral (fig. \ref{deduc})
that $H$ is also symmetric with respect to $(~,~)$:
\be\label{zillow} (\Psi,H\chi)=(H\Psi,\chi). \ee  
However, it is not true that $H$ is equal to its own transpose, as this formula suggests.  Rather, according to eqn. (\ref{zellme}), taking the transpose of $H$ is equivalent to reversing
the orientation of $Y$.     In other words, eqn. (\ref{zillow}) is valid if and only if $\Psi$ and $\chi$ are defined with opposite orientations on $Y$.  Concretely,
given this assumption, the proof of eqn. (\ref{zillow}) is made by integrating by parts in the integral (\ref{dedo}) that defines the bilinear pairing $(~,~)$.  By contrast, in eqn. (\ref{limbo}),
$\Psi$ and $\chi$ are defined with the same orientation on $Y$.
 The two statements
(\ref{limbo}) and (\ref{zillow}) are equivalent, given that $H$ commutes with $\cT$.

In section \ref{secfour}, a different operator will play the role of $\cT$ in relating the two inner products. So it will be useful to know, starting with a bose-symmetric
bilinear pairing $(~,~)$, what properties the antilinear operator $\cT$ must have so that eqn. (\ref{lombo}) defines a hermitian pairing.   A hermitian pairing is supposed to satisfy
\be\label{zelbox}\la \Psi,\chi\ra =\overline{\la\chi,\Psi\ra}. \ee
If $\la~,~\ra$ is defined as in eqn. (\ref{lombo}), then this is equivalent to
\be\label{telbo} (\cT\Psi,\chi) =\overline{(\cT\chi,\Psi)}. \ee
Because of the symmetry of $(~,~)$, the right hand side also equals $\overline{(\Psi,\cT\chi)}$, so $\cT$ is the antilinear analog of a symmetric operator.

\section{Background on Discrete Spacetime Symmetries}\label{background}

Before trying to generalize the preceding analysis to theories  that do have reflection and time-reversal symmetries,
we  pause to discuss some general issues related to such symmetries. 
First of all, in four-dimensional Minkowski space, parity or $\sP$ is usually defined  to act, in some Lorentz
frame,  by a reflection of all three spatial coordinates, leaving the time fixed. Thus $\sP$ maps $(t,x,y,z)$ to $(t,-x,-y,-z)$.  On the other hand, time-reversal  $\sT$ is defined to  map $t$ to $-t$ while leaving fixed the spatial coordinates.  So the product
$\sP \sT$ acts on spacetime by $(t,x,y,z)\to (-t,-x,-y,-z)$.

If a Lorentz-invariant theory is continued to Euclidean signature by setting $t=-\i t_\sE$, then the resulting Euclidean theory automatically has a symmetry \be\label{delfus}(t_\sE,x,y,z)\to
(-t_\sE,-x,-y,-z),\ee  because this is simply the product of a $\pi$ rotation of the $t_\sE-x$ plane with a $\pi$ rotation of the $y-z$ plane.   The continuation of this symmetry back to Lorentz
signature gives the universal symmetry that is usually called $\CPT$.  
Like other symmetries that reverse the sign of the time, $\CPT$ is represented by an antilinear rather than linear operator in quantum mechanics.  It is not immediately apparent
why this is the case; see Appendix \ref{confusion} for an explanation.
 The reason that this symmetry is traditionally called $\CPT$, rather than $\sP\sT$, is that it anticommutes with all additive conserved or nearly conserved charges
(such as baryon number, lepton number, and electric charge in the real world).  To see why this is the case, observe that, as a rotation commutes with the symmetries generated by 
conserved charges, the Euclidean operation $(t_\sE,x,y,z)\to
(-t_\sE,-x,-y,-z)$ commutes with those symmetries, and its continuation to Lorentz signature therefore also commutes with the symmetry group generated by the conserved charges.
A typical element of that symmetry group is $e^{\i \alpha K}$, where $\alpha$ is a real parameter and $K$ is a real linear combination of hermitian conserved charges.  As the symmetry
that acts by $(t,x,y,z)\to (-t,-x,-y,-z)$  reverses
the direction of time, it is antilinear and anticommutes with $\i$.  To commute with $e^{\i \alpha K}$, it must then also anticommute with $K$, which is why it is traditionally called  $\CPT$.

For a general value of the  spacetime dimension $D$, the formulation in terms of $\CPT $ is inconvenient.   Indeed, in $D$ dimensional Euclidean space,
the operation \be\label{whats}(t_\sE,x_1,x_2,\cdots, x_{D-1})\to (-t_\sE,-x_1,-x_2,\cdots, -x_{D-1})\ee
 is contained in the connected component of the rotation group, making it an automatic symmetry, if and only
if $D$ is even.   To define a $\CPT$-like symmetry in a way that is valid for any $D$, it is convenient to consider instead of $\sP$ an operation $\sR$ that reflects
just one spatial coordinate.\footnote{Some authors have simply redefined $\sP$ to be a reflection of one coordinate, but this unfortunately clashes with standard terminology in particle
physics.}   The operation $(t_\sE,x_1,x_2,\cdots, x_d)\to (-t_\sE,-x_1,x_2,\cdots, x_d)$ is a $\pi$ rotation of the $t_\sE-x_1 $ plane, so it is a symmetry in any
rotation-invariant theory, regardless of $D$.    This symmetry continues in Lorentz signature to a universal symmetry that we may call $\CRT$.  Clearly, in contrast to $\CPT$,
$\CRT$ can be defined for any value of
$D$.   In section \ref{incfermions}, we explain a further (minor) simplification of $\CRT$ relative to $\CPT$.
 
 In ordinary quantum field theory without gravity (and in  topological field theory), one has a precise  theorem stating that any Lorentz-invariant quantum field theory has the universal symmetry that we are calling $\CRT$.   In gravity, there is no such  precise theorem, but we do expect that a general gravitational theory does have  $\CRT$ symmetry.  This is true in semiclassical theories
 of gravity, it is true in perturbative string theory,  it is true nonperturbatively in the AdS/CFT correspondence, and it is true in the framework of Euclidean path integrals considered
 in the present article. 
  
$\CRT$ is the only discrete spacetime symmetry that is universally expected in any theory, but many interesting theories have reflection and time-reversal
symmetries as well.   In the presence of such symmetries, there are several additional issues to consider.   Before being general, let us recall that in the real world there are several
additive conserved charges -- notably electric charge, baryon number, and lepton number -- that are exactly or at least very nearly conserved.  The approximate discrete spacetime
symmetries in the real world 
either commute or anticommute with these additive discrete charges.   Therefore, the possible approximate time-reversal symmetries are $\sT$ and $\sC\sT$,
where $\sT$ commutes with the conserved charges and $\sC\sT$ anticommutes with them,
 and similarly the possible approximate reflection symmetries are $\sR$ and $\sC \sR$.  These are the only approximate discrete spacetime symmetries in the real world, but in a general
 theory there are more possibilities.
  A   time-reversal or reflection symmetry might commute with some conserved charges and
 anticommute with others, and it might not be possible to fully characterize a time-reversal or reflection symmetry by saying how it transforms the discrete charges; it may also be necessary
 to specify the action of the symmetry on fields of the theory.   Thus the usual terminology used in particle physics is not adequate to discuss a general case.
 
 Instead of  complicating our notation to fully specify  how a given reflection symmetry acts on conserved charges and other fields,
 we will simplify the notation and just write $\sR$ for a reflection symmetry, and $\sT$ for the corresponding time-reversal symmetry, defined so that the product $\sR\sT$ is the universal
 symmetry associated to a $\pi$ rotation in Euclidean signature.   Thus, we will not specify in the notation how $\sR$ or $\sT$ acts on conserved charges or other fields
 and we will drop $\sf C$ from the notation.
 
 If a given theory does have a reflection symmetry $\sR$, it is not necessarily unique.   If  $\sR$ is a reflection symmetry, and $\rm W$ is an ``internal'' symmetry, not acting on spacetime, 
 then $\sR'=\sR{\rm W}$ is another reflection symmetry, albeit one that acts differently on the fields (and possibly on the conserved charges, if any are present).
 If we replace $\sR$ with $\sR'$, we can compensate by replacing $\sT$ with $\sT'={\rm W}^{-1}\sT$.   This will ensure that $\sR'\sT'=\sR\sT$ is the universal symmetry
 associated to a $\pi$ rotation in Euclidean signature.
 
 Reflection symmetries that satisfy $\sR^2=1$ are of particular importance, for the following reason.  Let us discuss the matter in Euclidean signature.   First of all, any relativistic
 theory in Euclidean signature has at least ${\sf SO}(D)$ symmetry.   If a theory also has a reflection symmetry that satisfies $\sR^2=1$, then by adjoining this to ${\sf SO}(D)$
 one gets an action of ${\sf O}(D)$ on the fields of that theory.  The holonomy group of a possibly unorientable $D$-manifold $X$ is ${\sf O}(D)$ or a subgroup thereof. 
  Thus an action  of ${\sf O}(D)$ on the fields of a theory is what we need  to define that theory on a general, possibly unorientable $D$-manifold $X$.  In parallel transport around
 a closed loop $\gamma\subset X$, fields are transformed by the appropriate holonomy element in ${\sf O}(D)$, which will be an element of ${\sf SO}(D)$ or of the non-identity component of
 ${\sf O}(D)$ 
 depending on whether the orientation of $X$ is preserved or reversed in going around $\gamma$.

 If a reflection symmetry with $\sR^2=1$ exists, it is not necessarily unique.   To illustrate this point, consider a theory that has a reflection symmetry that satisfies $\sR^2=1$,
 generating a group that we will call $\Z_2^\sR$, and also an ``internal'' symmetry $\Z_4$, generated by an operator $ \rm U$ that satisfies  ${\rm U} ^4=1$.   In ordinary quantum field theory and
 in topological field theory, $\Z_4$ might be a global symmetry or a gauge symmetry; in gravity, we expect that $\Z_4$ will be a gauge symmetry \cite{BanksSeiberg,HO,emergence}.   In a theory with the combined
 $\Z_2^\sR\times \Z_4$ symmetry, there are actually two reflection symmetries that square to 1, namely $\sR$ and $\sR {\rm U}^2$.   The same theory also has reflection symmetries
 whose square is not 1, namely $\sR \rm U$ and $\sR {\rm U}^3$.    
 
 In a variant of this example, there is no reflection symmetry that satisfies $\sR^2=1$.  For this, it suffices to explicitly break the group $\Z_2^\sR\times \Z_4$ to the group $\Z_4'$ generated
 by $\sR {\rm U}$.   To do this, if  $\Z_4$ is a global symmetry, we add to the action an operator that is invariant under $\Z_4'$ but not under any other elements of
 $\Z_2^\sR\times \Z_4$.   This will give a theory that has the reflection symmetries $\sR{\rm U}$ and $(\sR\rm U)^3=\sR {\rm U}^3$, but has no reflection symmetry whose square is 1.  
 The $\Z_4'$ symmetry can be gauged if we wish (assuming it is anomaly-free).   A reflection symmetric theory that has no reflection symmetry whose square is 1 cannot be defined on a generic unorientable manifold.  Such a theory can be defined
 on unorientable manifolds that possess a certain slightly exotic differential geometric structure.\footnote{\label{dolbo} In general, consider any group $G$ with a non-trivial
 homomorphism $\rho:G\to \Z_2^{\sR}$.
 Then $G$ can act as follows on $SO(D)$: elements of $G$ that map to the identity in $\Z_2^{\sR}$ act trivially, and the others act by a reflection.   This action defines a group
 $SO(D)\rtimes G$, endowed with a homomorphism $1\times \rho: SO(D)\rtimes G\to O(D)$.    An $SO(D)\rtimes G$ structure on a $D$-manifold $X$ is an $SO(D)\rtimes G$ bundle
 over $X$ that is mapped by $1\times \rho$ to the $O(D)$ bundle related to the tangent bundle of $X$.   A theory with $G$ symmetry can be defined on a possibly unorientable
manifold endowed with an $SO(D)\rtimes G$ structure.  In the example $G=\Z_4'$ discussed in the text, an unorientable manifold that does not have an $SO(D)\rtimes G$ structure
is ${\mathbb {RP}}^2$; one that does is $(S^2\times S^3)/\Z_4$, where $\Z_4$ acts freely on $S^3$ and a generator of $\Z_4$ acts on $S^2$ as an inversion $(x,y,z)\to (-x,-y,-z)$.}
 
In our further discussion of reflection-invariant theories, rather than exploring the more exotic possibilities mentioned in footnote \ref{dolbo}, we assume that  the reflection operator $\sR$ can be chosen to satisfy $\sR^2=1$.   As we explained in the example of $\Z_2^\sR\times \Z_4$, such a choice of $\sR$ is in general not unique.  Does the choice
matter?    If $\sR$ and $\sR'$
 are two reflection symmetries, then ${\rm W}=\sR\sR'$ is an ``internal'' symmetry (acting trivially on spacetime).   If ${\rm W}$ is a global symmetry, then  it does matter
 whether we use $\sR$ or $\sR'$ to define the theory on an unorientable manifold.\footnote{In some  cases, there may be a symmetry that conjugates $\sR$ into $\sR'$ and then
 modulo this symmetry the choice between them does not matter.}  The different choices can give genuinely inequivalent ways to define the same theory on an unorientable manifold.
   If ${\rm W}$ is a gauge symmetry, the choice between  $\sR$ and $\sR'$ does not matter, since in any computation one will anyway be summing over all possible
 twists by ${\rm W}$, and the difference between using $\sR$ or $\sR'$ to define the theory on an unorientable manifold is exactly such a twist.     Of course, in gravity, one expects that ${\rm W}$ will always be a gauge symmetry, so that the choice of $\sR$ will not matter.    In what follows, we assume that, 
 regardless of whether this choice is essentially unique or inequivalent choices would have been possible,
  we are given a specific reflection symmetry $\sR$ whose square is 1.  
 
 Going back to Lorentz signature, we now want to discuss time-reversal.   A choice of $\sR$ determines a choice of $\sT$, such that the product
 is the universal symmetry $\sR\sT$ related to a $\pi$ rotation in Euclidean signature.   In the absence of fermions, $\sR$ and $\sT$ commute, because they
 arise in Euclidean signature from reflections in orthogonal directions  ($\sT$ is a reflection of Euclidean time, and $\sR$ is a reflection of a Euclidean space direction), which commute.  Fermions will be included in section \ref{incfermions}; for now, we assume that $\sR$ and $\sT$ commute.
 In the absence of fermions, $(\sR\sT)^2=1$, since in Euclidean signature $\sR\sT$  is just a $\pi $ rotation.  With $\sR^2=(\sR\sT)^2=1$ and $\sT$ commuting with $\sR$, we have
 $\sT^2=1$.   This then defines the class of theories that we will consider in section \ref{secfour}: theories with chosen reflection and time-reversal symmetries $\sR$ and $\sT$
such that $\sR\sT$ is the universal symmetry related to a rotation in Euclidean signature and 
 \be\label{ximbo}\sR^2=\sT^2=1,~~\sR\sT=\sT\sR.  \ee
 
In Minkowski space, $\sR$ and $\sT$ play somewhat similar roles as discrete spacetime symmetries, though they differ in that $\sR$ is unitary and $\sT$ is antiunitary.
When we quantize a reflection-invariant  theory on a general $D-1$-manifold $Y$ to get a Hilbert space $\H_Y$, $\sR$ and $\sT$  play very different roles.   There is in general nothing that one could call reflection symmetry
acting on  $\H_Y$; such symmetries exist to the extent that $Y$ admits an orientation-reversing isometry.   For general $Y$, there is no such isometry (and for some $Y$'s,
there are many inequivalent ones so there are inequivalent $\sR$-like symmetries).  Rather,
the role of $\sR$ is that, as explained earlier, 
it enables us to define $\H_Y$ even if $Y$ is unorientable (and without choosing an orientation of $Y$ if $Y$ is orientable).
By contrast, for any $Y$, $\sT$ is an antilinear symmetry of $\H_Y$. This statement, however, is not quite trivial:
it  depends on the fact that $\sR$ and $\sT$ commute.
If $\sT$ and $\sR$ did not commute, we would  have $\sT^{-1}\sR\sT=\sR'$, where $\sR'$ is some other reflection symmetry.   In that case, acting with $\sT$ would
map $\H_Y^{\sR}$ to $\H_Y^{\sR'}$, where $\H_Y^{\sR}$ and $\H_Y^{\sR'}$ are the Hilbert spaces of $Y$ defined  using respectively $\sR$ or $\sR'$.  Something like
this actually happens in the presence of fermions, as explained in section \ref{incfermions}.   But for now we assume that $\sR$ and $\sT$ commute, in which case $\sT$
can always be defined as an antilinear symmetry of $\H_Y$.     In a theory of gravity,  for the same reasons as in the discussion of $\cT$ in section \ref{orhilb}, it seems that we should
not consider $\sT$-invariant states to be the only physically sensible ones.

Since $\sT$ is a natural symmetry of any $\H_Y$ and $\sR$ is not, it follows that
also there is no natural $\sR\sT$ symmetry of $\H_Y$.  Any such symmetry depends on the existence and choice of an orientation-reversing symmetry of $Y$.
In the case of gravity in a closed universe, if $Y$ does have an orientation-reversing symmetry $\rho$, then $\rho$ will behave as a gauge symmetry (fig. \ref{bucnot}).

\section{Theories With A Reflection Symmetry}\label{secfour}

We now consider a theory with chosen  $\sR$ and $\sT$ symmetries that satisfy the conditions of eqn. (\ref{ximbo}).   In such a theory, we want to understand the relation
between the permutation symmetric pairing $(~,~)$ that is naturally defined by the path integral, and the positive-definite hermitian pairing $\la~,~\ra$ that determines probabilities in  quantum mechanics.
In analyzing this question in section \ref{sectwo} for theories that lack reflection and time-reversal symmetry, a key role was played by the fact that the Hilbert space $\H_Y$ defined in quantizing
such a theory on a $D-1$-manifold $Y$ depends on a choice of orientation of $Y$.   A reflection-symmetric theory
 can be defined on a manifold $Y$ even if $Y$ is unorientable,  so we cannot make use of a choice of orientation.  
Therefore, the operator $\cT$ considered in section \ref{sectwo}, which acts by reversing the orientation of $Y$ and complex conjugating the wavefunction, is not available.

In general, the only antilinear symmetry that is available is $\sT$.  So  the relation between the two pairings will have to be
\be\label{zingo} \la\Psi,\chi\ra=(\sT\Psi,\chi). \ee  Since $\sT^2=1$, we also have
\be\label{wingo}\la \sT\Psi,\chi\ra= (\Psi,\chi). \ee
Of course, $\la~,~\ra$ and $(~,~)$ coincide if we restrict to $\sT$-invariant states.   

Let us see how this works in a concrete example of a theory with $\sR$ and $\sT$ symmetry.  An example in which $\sT$ acts merely by complex-conjugating the wavefunction,
as in Einstein gravity or  gauge theory without a theta-angle, is too simple to be representative.   As a more representative example, we return to the four-dimensional gauge theory with a theta-angle that was studied in section \ref{sectwo},
but now we restore reflection and time-reversal symmetry by replacing the parameter $\theta$ with an axion-like field $a$.   Thus the action is obtained by replacing $\theta$ with
$a$ in eqn. (\ref{lorac}):
\be\label{zorack}I_L =-\frac{1}{4e^2}\int \d^4x \sqrt g\, \tr\,F_{\mu\nu} F^{\mu\nu} +\frac{1}{16\pi^2}\int\d^4x\sqrt{g} \,a \epsilon^{\mu\nu\alpha\beta}\tr\,F_{\mu\nu}F_{\alpha\beta}.\ee
(We omit additional $\sR$ and $\sT$ invariant terms, such as the kinetic energy of $a$, that are needed to fully specify a model but do not play a role in the following.)
The Hamiltonian, again with some terms omitted, is obtained from eqn. (\ref{amton}) by the same substitution:   
\be\label{zamton}H=\int_Y\d^3x \sqrt h\,\tr \left(\frac{e^2}{2}\left(-\i\frac{\delta}{\delta A_i(x) }-\frac{a}{8\pi^2} \t\epsilon_{ijk}F^{jk}(x)\right)^2+\frac{1}{4e^2} \tr\,F_{ij}F^{ij}\right) .\ee
This theory has $\sR$ and $\sT$ symmetry, with $a$ being odd under $\sR$ and $\sT$; for example, if $t$ and $\vec x$ parametrize time and space in Minkowski space,
then $\sT$ maps $a(t,\vec x)$ to $-a(-t,\vec x)$.    

A quantum wavefunction in this model is a complex-valued function $\Psi(A,a)$. 
Time-reversal acts by complex conjugation of the wavefunction and a sign change of $a$:
\be\label{dobbo|}\sT\Psi(A,a)=\bar\Psi(A,-a).\ee
  The hermitian inner product in $\H_Y$ that computes quantum mechanical probabilities is formally defined by an integral over fields on $Y$:
\be\label{zintegral} \la\Psi,\chi\ra=\frac{1}{{\rm vol}\,\mathcal G}\int DA\,Da \,\bar\Psi(A,a)\chi(A,a).\ee 
Given the action of $\sT$, the bose symmetric pairing that comes from the path integral is
\be\label{omingo} (\Psi,\chi)=\frac{1}{{\rm vol}\,\mathcal G}\int DA\,Da \,\Psi(A,a)\chi(A,-a).\ee

What might be unexpected in this formula is that $\Psi$ depends on $a$ while $\chi$ depends on $-a$.
That is actually consistent with permutation symmetry, $(\Psi,\chi)=(\chi,\Psi)$, since the definition of $(~,~)$ involves an integration over all $a$.
However,  the relative minus sign may be unexpected, so we will give some further explanations of why it must be present.

First, as in fig. \ref{deduc}, the Hamiltonian $H$ is supposed to be symmetric with  respect to $(~,~)$.   However, a look at eqn. (\ref{zamton}) shows that the transpose of $H$ is
an operator of the same kind with $a$ replaced by $-a$.   So the minus sign in the definition of the bose symmetric inner product is needed to ensure that
\be\label{kinzo}(\Psi,H\chi)=(H\Psi,\chi), \ee
as formally predicted by the path integral.  At the same time, $H$ is hermitian, so it also satisfies
\be\label{inzo}\la \Psi,H \chi\ra =\la H\Psi,\chi\ra.\ee
These statements are equivalent to each other, as $H$ commutes with $\sT$.  

However, one would also like a more microscopic explanation of  the minus sign in the inner product.   For this, we return to the action (\ref{zorack}) and ask how a theory with 
this action can be defined on an unorientable four-manifold $X$.   Clearly, $a$ cannot be regarded as an ordinary scalar field; it must change sign in going around any orientation-reversing
loop in $X$.  
The antisymmetric tensor $\epsilon^{\mu\nu\alpha\beta}$ that appears in the action has the same property.   What is well-defined and single-valued is the 
product $a_{\mu\nu\alpha\beta}=a\epsilon_{\mu\nu\alpha\beta}$ which actually appears in the action.   The meaning of the action (\ref{zorack}) is therefore more transparent if 
we express the action in terms of the four-form $a_{\mu\nu\alpha\beta}$ rather than in terms of a scalar field $a$.   The action is then manifestly well-defined since only well-defined,
single-valued quantities appear in it:
\be\label{yorac}I_L =-\frac{1}{4e^2}\int \d^4x \sqrt g\, \tr\,F_{\mu\nu} F^{\mu\nu} +\frac{1}{16\pi^2}\int\d^4x\sqrt{g} a^{\mu\nu\alpha\beta}\tr\,F_{\mu\nu}F_{\alpha\beta}.\ee
Similarly, the Hamiltonian (\ref{zamton}) depends on the field $a$ and the three-dimensional Levi-Civita tensor $\t \epsilon_{ijk}$, both of which change sign in going around
an orientation-reversing loop in $Y$.   However, the product $\t a_{ijk}=a\t\epsilon_{ijk}$ which actually appears in the Hamiltonian is single-valued, so we can write a manifestly well-defined formula for the Hamiltonian 
by simply expressing it in terms of a three-form $\t a_{ijk}$ on $Y$ rather than a double-valued scalar field $a$:
\be\label{yamton}H=\int_Y\d^3x \sqrt h\,\tr \left(\frac{e^2}{2}\left(-\i\frac{\delta}{\delta A_i(x) }-\frac{1}{8\pi^2} \t a_{ijk}F^{jk}(x)\right)^2+\frac{1}{4e^2} \tr\,F_{ij}F^{ij}\right) .\ee

With this understood, let us return to the path integral of fig. \ref{one}(a) on a manifold $X$ that formally defines a permutation symmetric coupling among states of closed universes $Y_r$, $r=1,\cdots,s$.
On $X$, the action is defined in terms of a four-form field $a_{\mu\nu\alpha\beta}$.   On $Y_r$, the Hamiltonian is defined in terms of a three-form $\t a_{ijk}$.
Somewhat as in section \ref{hermin}, the relation between $a_{\mu\nu\alpha\beta}$ in bulk and $\t a_{ijk}$ on the $r^{th}$ boundary component can be formulated as follows.
Near each $Y_r$, pick a Euclidean time coordinate $x^0=t_r$ that is constant on $Y_r$ and increases away from $Y_r$, and such that the metric of $X$ near $Y_r$ is
$\d t_r^2+h_{ij}\d x^i\d x^j$.    Then along $Y_r$, we define $\t a_{ijk}=a_{0ijk}$.   Following this recipe for every boundary component will give the analog of what
we had in section \ref{hermin} in discussing theories without time-reversal or reflection symmetry. 

Let us specialize to the case relevant to the bilinear pairing $(~,~)$. 
 For this, we take $X=Y\times I$, with a metric of the form $\d t^2+h_{ij} \d x^i\d x^j$, where $t\in [0,\tau]$ parametrizes the interval $I$, and $h_{ij}\d x^i\d x^j$ is a metric on $Y$.
To compute the pairing, we take a limit of small $\tau$, and the bulk field $a_{\mu\nu\alpha\beta}$ becomes independent of $t$.   The three-forms $\t a^{(0)}_{ijk}$ and
$\t a^{(\tau)}_{ijk}$ at $t=0$ and $t=\tau$ are defined by 
\be\label{modo}\t a^{(0)}_{ijk}=a_{0ijk},~~~~\t a^{(1)}_{ijk}=-a_{0ijk}. \ee
The reason for the minus sign in the second formula is that $t$ increases away from the boundary of $X$ at $t=0$, but decreases away from the boundary at $t=\eta$.  
So $\t a^{(0)}_{ijk}=- \t a^{(1)}_{ijk}$, which accounts for the relative minus sign in the formula (\ref{omingo}) for the pairing.

\section{Including Fermions}\label{incfermions}

\subsection{Fermions and Discrete Spacetime Symmetries}\label{fermsym}

Including fermions requires some small but important modifications of our previous statements.   The main reason for this is that any theory with fermions has
the universal symmetry $(-1)^{\sF}$ that equals 1 or $-1$ for bosonic or fermionic states.  This symmetry enters in many statements.

First we will describe the action on fermions of the universal symmetry $\CRT$.   
This action is only uniquely determined up to an overall sign because we are free to replace $\CRT$
by $\CRT (-1)^{\sF}$.  However, the arbitrary sign is the same for all fermions.
In analyzing the action of $\CRT$, it suffices to consider the case of a real fermion field $\lambda$ (possibly of definite chirality if the dimension $D$ is congruent to 2 mod 4).
Complex fermion fields need not be considered separately, because a complex fermion 
field $\Lambda$  can be expanded  $\Lambda=\lambda_1+\i\lambda_2$ in terms of a pair of real fermion fields.    Real gamma matrices,
satisfying a Clifford algebra 
\be\label{inhu} \{\Gamma_\mu,\Gamma_\nu\}=\pm 2\eta_{\mu\nu},\ee can always be chosen so that the Dirac equation 
$\Gamma^\mu\partial_\mu\lambda=0$ is real.  For some values of $D$, in finding real gamma matrices of minimum dimension 
we have to take a $+$ sign in the Clifford algebra, for some values of $D$
we need a $-$ sign, and for some values of $D$, both signs are possible.   The minimum dimension of the space on which the Clifford
algebra can be represented by real matrices depends on $D$, but whenever a real fermion 
field $\lambda$ can exist in a relativistic field theory, there is a corresponding set of real gamma matrices.\footnote{A particular real fermion
field $\lambda$ might have a bare mass or other couplings, but it is always possible to write a real massless Dirac equation for 
$\lambda$.  It may happen (for some values of $D$, such as $D=4$) that
$\lambda$ transforms in a representation of the spin group ${\rm SO}(1,D-1)$ that is irreducible as a real representation but 
reducible as a complex representation.  That is not relevant for our present
purposes.  If $D=4k+2$, $\lambda$ can satisfy a chirality condition; this will also not affect the discussion of $\CRT$, because the chirality operator
$\Gamma_0\Gamma_1\cdots \Gamma_{D-1}$ commutes with $\Gamma_0\Gamma_1$.}  In a moment we will see that in a certain sense the sign
in the Clifford algebra does not matter.
The transformation of a real fermion field $\lambda$ under $\CRT$ is  
\be\label{zono}\CRT\lambda(t,x_1,x_2,\cdots,x_d)\CRT^{-1}=\pm \Gamma_0\Gamma_1 \lambda(-t,-x_1,x_2,\cdots,x_d).\ee   We include  the arbitrary sign
mentioned earlier.  This sign is the same for all fermions, since otherwise $\CRT$ symmetry could not be universal: by 
adding to the Hamiltonian terms that are bilinear in fermions that transform with
opposite signs, we could violate $\CRT$.  We note that (\ref{zono}) is indeed a symmetry of the massless Dirac equation $\Gamma^\mu\partial_\mu\lambda=0$, 
and it remains a symmetry if a mass term is added (in cases in which this is possible).  
   Regardless of the signs in eqns.  (\ref{inhu}) and (\ref{zono}), we always have $(\pm \Gamma_0\Gamma_1)^2=1$, and therefore for any $D$, we always have $\CRT^2=1=(\CRT (-1)^{\sF})^2$.   By contrast, for even $D$ where $\CPT$ can be defined, $\CPT^2$ is equal to 1 or $(-1)^{\sF}$, depending on $D$.  So in this respect,
the properties of $\CRT$ are more uniform.

As a check on the conclusion that $\CRT^2=1$, 
 consider the example of a one component Majorana-Weyl fermion $\lambda(x,t)$ in two spacetime dimensions.  As $\lambda$  has just one hermitian component, any conceivable
  $\CRT$ operator satisfies $\CRT^2=1$ (to get $\CRT^2=(-1)^\sF$ one needs an even number of hermitian fermion components).   The following may be puzzling. 
   $\CRT$ is   related to a $\pi$ rotation in Euclidean signature, which multiplies $\lambda$ by $\i$ (or $-\i$, depending on conventions). So the square of a $\pi$ rotation equals $(-1)^\sF$,
   in contrast to the fact that $\CRT^2=1$.  Actually, all this is consistent with the fact that
   the Lorentz signature two-point function $\la\Omega|\lambda(t_1,x_1)\lambda(t_2,x_2)|\Omega\ra$ is an analytic continuation of the Euclidean two-point function
   $\la \lambda(t_{\sE,1},x_1)\lambda(t_{\sE,2},x_2)\ra$ via $t=\i t_\sE$.   In Euclidean signature, under a $\pi$ rotation, the correlation function is multiplied by $\i^2=-1$.  In Lorentz signature,
   there are no such factors of $\i$.   But because $\CRT$ is antiunitary and self-adjoint,  it reverses the order of multiplication of the two fermion operators,\footnote{An antilinear selfadjoint operator $\Theta$ satisfies $\la\Psi|\Theta\chi\ra
   =\overline{\la\Theta\Psi|\chi\ra}=\la\chi|\Theta\Psi\ra$.   For operators $A,B$, write $A'=\Theta\A\Theta^{-1}$,  $B'=\Theta B\Theta^{-1}$.  If $\Theta^2=1$, $\Theta|\Omega\ra=|\Omega\ra$,
   then $\la\Omega|A B|\Omega\ra = \la \Theta\Omega|A B|\Theta\Omega\ra =\overline{\la\Omega|\Theta AB\Theta|\Omega\ra}=\overline{\la\Omega|A'B'\Omega\ra}=\la A'B'\Omega|\Omega\ra =\la\Omega|B'^\dagger A'^\dagger|\Omega\ra$.  In the case at hand with $\Theta=\CRT$, we have
    $\lambda(t,x)'^\dagger=\lambda(-t,-x)^\dagger=\lambda(-t,-x)$, so the effect of $t,x\to -t,-x$ is merely to reverse the order in which the two fermion operators
   are multiplied.} giving a factor of $-1$ under $(x,t)\to (-x,-t)$, as expected
   based on analytic continuation from Euclidean signature.   For a correlation
   function with $n=2k$ fermion operators, in Euclidean signature one gets a factor $\i^n=(-1)^k$, and in Lorentz signature one gets the same factor from reversing the order of a product of $n$
   fermion operators.   The minus sign associated to fermi statistics that is needed here for consistency shows the close relation of the spin-statistics theorem to the $\CRT$ theorem.\footnote{The way that the antilinear nature of $\CRT$ interacts with the ordering of operators is also important in understanding why a $\pi$ rotation in Euclidean signature becomes an antilinear operator
   on Hilbert space.   See Appendix \ref{confusion}.}

In contrast to $\CRT$, whose action on fermions is universal, separate reflection and time-reversal symmetries act on fermions in a way that is highly
non-universal, in general.  As in section \ref{background},  general reflection and time-reversal symmetries cannot be labeled simply by whether they do or do not act via $\sC$.
Given this, we will as in section \ref{background} drop $\sC$ from the notation.   Thus we refer to the universal symmetry as simply $\sR\sT$, and if reflection and time-reversal
symmetries are present, we denote them simply as $\sR$ and $\sT$, without specifying how they transform discrete symmetries or other fields.

With this understood, let us discuss a general choice of $\sR$ for a theory with fermions.    For any choice of $\sR$, we then choose $\sT$ so that $\sR\sT$ is the universal
symmetry that acts on fermions as in eqn. (\ref{zono}).   In any dimension $D$, in a reflection-symmetric theory, spin 1/2 fermions transform under the Lorentz group as
the direct sum of $n$ copies $\lambda_i$, $i=1,\cdots , n$  of a certain
real irreducible representation of the Clifford algebra.\footnote{In the absence of reflection symmetry,
this is not true: in $4k+2$ dimensions, the Lorentz group has  inequivalent real spinor
representations of positive and negative chirality.  For odd $D$, although the real irreducible  spinor representation of
the Lorentz group is unique, there are two corresponding representations of the Clifford algebra, differing by $\Gamma_i\to -\Gamma_i$.   For our purposes, this sign
is inessential as it can be absorbed in the matrix $M$ of eqn. (\ref{genac}).}  
A rather general  action of $\sR$ on $n$ such multiplets, giving a symmetry of the massless Dirac equation, is then
\be\label{genac}\sR\lambda_i(t,x_1,x_2,\cdots, x_d)\sR^{-1}=\Gamma_1\sum_j M_{ij}  \lambda_i(t,-x_1,x_2,\cdots, x_d),\ee
where $M$ is an $n\times n$ orthogonal matrix.  This action of $\sR$  can be generalized for some even values of $D$ by
making use of the chirality operator $\bar\Gamma=\Gamma_0\Gamma_1\cdots\Gamma_d$.   However, (\ref{genac}) will suffice for illustration.

In a theory containing fermions, the simple realizations of reflection symmetry are such that $\sR^2=1$ or $\sR^2=(-1)^\sF$.   (This generalizes the fact
that in a theory of bosons only, simple realizations of reflection symmetry are such that $\sR^2=1$.)  Of course, to get $\sR^2=1$ or $(-1)^{\sF}$ involves
a restriction in how $\sR$ acts on bosonic fields as well as a condition on its action on fermions, but the constraints on the action on fermions are easily
stated.     To get $\sR^2=1$, we can have $M^2=1$ and a $+$ sign
in the Clifford algebra (\ref{inhu}), or $M^2=-1$ and a $-$ sign in the Clifford algebra.   To get $\sR^2=(-1)^{\sF}$, which means that $\sR^2=-1$ on fermions, 
we can have $M^2=1$ and a $-$ sign in (\ref{inhu}), or $M^2=-1$
and a $+$ sign.  So both cases $\sR^2=1$ and $\sR^2=(-1)^{\sF}$ are possible for any $D$.

Once we pick $\sR$, $\sT$ is determined by the requirement that $\sR\sT$ is the universal symmetry that acts as in eqn. (\ref{zono}).
So
\be\label{benac}\sT\lambda_i(t,x_1,x_2,\cdots,s,x_d)\sT^{-1}=\mp\Gamma_0 \sum_j M^{-1}_{ij}\lambda_j(-t,x_1,x_2,\cdots,x_d).   \ee
We see that $\sR$ and $\sT$ do not commute; they anticommute in acting on fermions.   Since $\sR$ and $\sT$ commute in acting on bosons but anticommute on
fermions,
we have
\be\label{bilf} \sR\sT=(-1)^{\sF} \sT\sR. \ee    
For $M^2=\pm 1$, we have also
\be\label{ziff}\sR^2=\sT^2(-1)^{\sF},\ee
meaning that either $\sR^2=1$, $\sT^2=(-1)^{\sF}$ or $\sR^2=(-1)^{\sF}$, $\sT^2=1$.   Note that $\sR$ and $\sT$ map bosons to bosons and fermions
to fermions, so they commute with $(-1)^{\sF}$.   

\subsection{Comparing the Two Inner Products}\label{fermcomp}

Now we will discuss the relation between the hermitian inner product $\la~,~\ra$ and the bilinear form $(~,~)$.    First we consider  theories with no time-reversal or reflection symmetry,
and then theories that do have those symmetries.

As in section \ref{sectwo}, a theory without a reflection symmetry has to be formulated on a manifold $X$ that is not just orientable but has a chosen orientation
$\veps$.   The holonomy of the Riemannian connection around  a loop $\gamma\subset X$ is an element of $\SO(D)$.  This holonomy describes parallel
transport of tensors around $\gamma$.   If fermions are present, one needs a way to describe the parallel transport of a spinor around $\gamma$.  For this, one needs to 
be able to distinguish a trivial operation from a $2\pi$
rotation, which acts as $(-1)^{\sF}$.   A $2\pi$ rotation is a trivial operation in $\SO(D)$, but it is nontrivial in a double cover of $\SO(D)$ that is known as $\Spin(D)$:
\be\label{mitto}1\to \Z_2\to \Spin(D)\to \SO(D)\to 1. \ee
Here the non-trivial element of $\Z_2$ is $(-1)^{\sF}$.   A spin structure\footnote{For a more detailed description of spin structures and pin structures, see for example
Appendix A of \cite{Phases}.}  on $X$ is a $\Spin(D)$ bundle that has the property that if one projects it to $\SO(D)$,
one gets the $\SO(D)$ bundle related to the tangent bundle of $X$.   We denote a spin structure as $\s$.   A spin structure is needed to define fermions\footnote{Just as not every manifold
has an orientation, not every orientable manifold has a spin structure.   For example, there is no spin structure on $\Bbb{CP}^2$.} on $X$, since it enables us to define the holonomy in
parallel transport of a fermion field.

Similarly, to define a space of physical states for such a theory on a $D-1$-manifold $Y$, one requires on $Y$ a spin structure $\s_Y$, as well as an orientation $\veps_Y$.
In a theory with fermions and no reflection symmetry, one defines for each pair $\veps_Y,\s_Y$ a corresponding Hilbert space $\H_{\veps_Y,\s_Y}$.   For theories without
reflection symmetry, the presence of fermions does not significantly modify the relation between the hermitian form $\la ~,~\ra$ and the bilinear form $(~,~)$
that was described in section \ref{sectwo}.
 The operation
$\cT$ considered in section \ref{sectwo} that complex conjugates the wavefunction and reverses the sign of $\veps_Y$ can still be defined in the presence of fermions.
It leaves unchanged the holonomies that define  $\s_Y$, so it leaves  $\s_Y$ invariant (up to a natural isomorphism\footnote{\label{detailmath} Mathematically, the orientation of $X$ is built
into the definition of a spin structure on $X$, and one says that
 there is a natural isomorphism $\zeta$  from a spin structure $\s$ defined with one orientation to a spin structure $\zeta(\s)$ defined
with the opposite orientation; the isomorphism is such that parallel transport of a fermion around a general loop $\gamma\subset X$ gives the same result for $\s$ or for $\zeta(\s)$.
For our purposes we can just think of a spin structure as being defined by the fermion holonomies and $\zeta$ as the identity.}).  The hermitian pairing of quantum mechanics is a hermitian form $\la~,~\ra:\H_{\veps_Y,\s_Y}\otimes \H_{\veps_Y,\s_Y}\to \C$.
The bilinear pairing $(~,~)$ is related to $\la~,~\ra$ by $\la\Psi,\chi\ra=(\cT\Psi,\chi)$.  It is thus a pairing between $\H_{\veps_Y,\s_Y}$ and $\H_{-\veps_Y,\s_Y}$.  As in the absence
of fermions, the distinction between bras and kets in a Euclidean path integral is determined by the induced orientation of the boundary.

  If $X$ is unorientable, its Riemannian holonomy is
contained in a group $\O(D)$ that is generated by $\SO(D)$ together with a reflection $\sR$.   On an orientable manifold with fermions present, one has
to consider the holonomy as an element of $\Spin(D)$ rather than $\SO(D)$.   On an unorientable manifold with fermions present, one has to replace
$\Spin(D)$ with a group generated by $\Spin(D)$ together with a reflection.   This group is called $\Pin^+(D)$ if $\sR^2=1$ and $\Pin^-(D)$ if $\sR^2=(-1)^{\sF}$.
Note that    $(-1)^{\sF}$ is a central  element of $\Spin(D)$, so $\sR$ generates an outer automorphism of $\Spin(D)$ that is of order 2 whether $\sR^2$ equals 1 or $(-1)^\sF$.
Thus both $\Pin^+(D)$ and $\Pin^-(D)$ are, roughly, doubled versions of $\Spin(D)$.  They are both extensions of $\O(D)$ by a group $\Z_2$ generated by
$(-1)^{\sF}$:
\be\label{inoc} 1\to \Z_2\to \Pin^\pm(D) \to \O(D)\to 1. \ee

To define fermions on an unorientable manifold, using a reflection symmetry with $\sR^2=1$ or $\sR^2=(-1)^\sF$, requires what is known as a\footnote{Similarly to orientations
and spin structures, not every unorientable manifold has a pin$^+$ or pin$^-$ structure.  For example, $\Bbb{RP}^2$ has no pin$^+$ structure, $\Bbb{RP}^4$ has
no pin$^-$ structure, and ${\Bbb{RP}}^2\times {\Bbb{RP}}^2$ has neither type of structure.}  pin$^+$ structure  (if $\sR^2=1$)
or a pin$^-$ structure (if $\sR^2=(-1)^\sF$).    A pin$^+$ or pin$^-$ structure on a manifold $X$ is a $\Pin^+(D)$ or $\Pin^-(D)$ bundle over $X$  that projects in $\O(D)$ to the $\O(D)$ bundle related to the tangent bundle of $X$.
 So to define a reflection-invariant $D$-dimensional  theory with fermions,  one requires a pin structure $\p$
on spacetime  (more precisely, a pin$^+$ or pin$^-$ structure, as the case may be, but we will not indicate this distinction in the notation).   
Similarly, to quantize such a theory on a $D-1$-manifold $Y$ and define a space of physical states requires a pin structure $\p_Y$ on $Y$.
Given such a choice, one  can define a Hilbert space $\H_{Y,\p_Y}$.  In a theory with gravity, one would take the direct sum of $\H_{Y,\p_Y}$ over all $\p_Y$ to get
the Hilbert space associated to $Y$.  ($Y$ might have diffeomorphisms that permute the $\p_Y$'s and if so physical states are invariant under these diffeomorphisms.) 
	
For the limited purposes of the present article, there is one important difference between spin structures and pin structures.    In a theory with reflection symmetry, the closest 
analog of the operator $\cT$ that complex conjugates the wavefunction and reverses the orientation is time-reversal $\sT$.    $\cT$ leaves spin structures unchanged (up
to a natural isomorphism; see
footnote \ref{detailmath}), but because $\sR$ and $\sT$ do not commute,
 $\sT$ does not act trivially on pin structures.  

  To understand the  action of $\sT$ on pin structures, we need the notion of complementary pin structures.
  If $\p$ is a pin structure on a Riemannian manifold $W$, and $\gamma$ is a loop in $W$, let $\Hol_\p(\gamma)$ be the holonomy around $\gamma$ in the pin structure $\p$.
For every $\p$, there is a ``complementary'' pin structure $\p'$ such that $\Hol_{\p'}(\gamma)=\pm\Hol_\p(\gamma)$, where
the sign is $+$ if the loop $\gamma$ is orientation-preserving and $-$ if it is orientation-reversing.
 
The relation $\sR\sT=\sT\sR(-1)^{\sF}$ (eqn. (\ref{bilf})) can be read to say that  $\sT$ conjugates $\sR$ to $\sR (-1)^\sF$.  
If $\sR$ is replaced by $\sR (-1)^\sF$,  the holonomy around a loop $\gamma$ gets an extra minus sign for fermions if and only if this holonomy involves a reflection,
in other words if and only if $\gamma$ is orientation-reversing. This is precisely the operation that produces $\p'$ from $\p$.
 So $\sT$ maps the pin structure $\p$ to $\p'$, and therefore maps $\H_{Y,\p_Y}$ to
$\H_{Y,\p'_Y}$.   

The Hilbert space inner product $\la~,~\ra$, being positive definite, is a pairing of $\H_{Y,\p_Y}$ with itself, for any $Y$ and $\p_Y$.
Therefore the permutation symmetric inner product, satisfying as usual 
 $(\sT \Psi,\chi)=\la\Psi,\chi\ra$, pairs  $\H_{Y,\p_Y}$ with 
$\H_{Y,\p'_Y}$.   This is somewhat analogous to the observation made in section \ref{sectwo} that in a model with no reflection symmetry, $(~,~)$
pairs $\H_{Y,\veps_Y}$ with $\H_{Y,-\veps_Y}$.

Microscopically, the fact that $(~,~)$ pairs Hilbert spaces defined with complementary pin structures, not with the same pin structure,
can be demonstrated by an argument similar to the one used at the end of  section \ref{secfour} to  analyze the pairing $(~,~)$ in a particular reflection-symmetric theory. 

\appendix

\section{How Does Wick Rotation Turn A Linear Symmetry Into An Antilinear One?}\label{confusion}

In this appendix, we will discuss the following possibly confusing question.   Quantum field theory in Minkowski space has a universal $\CRT$ symmetry\footnote{As
explained in section \ref{background}, in four dimensions $\CRT$ symmetry is equivalent to the usual $\CPT$ symmetry.}  that acts on spacetime by
$(t,x_1,x_2,\cdots,x_{D-1})
\to (-t,-x_1,x_2,\cdots, x_{D-1})$.  Up to a certain point, Wick rotation to Euclidean signature by $t=\i t_\sE$ makes it obvious why this is true.   In Euclidean signature the
operation \be\label{bobbo}\P: (t_\sE,x_1,x_2,\cdots, x_{D-1})\to (-t_\sE,-x_1,x_2,\cdots, x_{D-1})\ee
 is simply a $\pi$ rotation of the $t_\sE-x_1$ plane, so it is a symmetry of any rotation-invariant theory.
After continuation back to Lorentz signature, this gives a Lorentz signature symmetry $(t,x_1,x_2,\cdots,x_{D-1})
\to (-t,-x_1,x_2,\cdots, x_{D-1})$, and this is $\CRT$.

What this does not explain is why $\CRT$ is antilinear.   The rotation symmetry $\P$ in Euclidean signature acts linearly on correlation functions, not antilinearly.
Under $\P$, an operator $\phi(t_\sE,x_1,\cdots, x_{D-1})$ transforms as follows:  the coordinates $(t_\sE,x_1,x_2,\cdots,x_{D-1})$ are rotated to  $(-t_\sE,-x_1,x_2,\cdots,x_{D-1})$,
and in addition, if $\phi$ carries spin, then $\P$ acts in some fashion on the spin labels of $\phi$.  We will just denote this action as $\phi\to \t \phi$, so overall $\P$
transforms 
\be\label{undu}\phi(t_\sE,x_1,x_2\cdots,x_{D-1})\to \t\phi(-t_\sE,-x_1,x_2\cdots,x_{D-1}).\ee   The implication of $\P$ symmetry for correlation functions is thus
\be\label{zorro} \la \phi_1(t_{\sE_1},x_{1,1},\cdots) \cdots \phi_d(t_{\sE,d},x_{1,d},\cdots)\ra = \la \t\phi_1(-t_{\sE,1},-x_{1,1},\cdots)\cdots \t\phi_d(-t_{\sE_d},-x_{1,d},\cdots)\ra. \ee
(For brevity we omit the coordinates on which $\P$ acts trivially.)
This is a linear relation between correlation functions, not an antilinear one, so why is it that when we interpret this relation as the consequence of the existence of 
a symmetry operator $\CRT$, this operator is antilinear?

To answer this question, we have to review how Euclidean correlation functions are given a Hilbert space interpretation.   To do this, one picks a direction in space which will be interpreted
as the Euclidean time.   Then one  introduces a Hilbert space of states at fixed Euclidean time, a vacuum state $|\Omega\ra$ in this Hilbert space, and a transfer  matrix
$e^{-\tau H}$ that 
propagates states in Euclidean time by an amount  $\tau$. The operator $H$ is interpreted as the Hamiltonian and is bounded below but not above; accordingly,
 $e^{-\tau H}$ makes sense as a Hilbert space operator only for
non-negative $\tau$.   Because of rotation symmetry, one can pick any direction at all in Euclidean space $\R^D$ and view it as the Euclidean time direction in developing
a Hilbert space interpretation of the theory.   However, if one wishes to address the question of why the specific rotation operator $\P$ becomes an antilinear symmetry in Lorentz signature,
then we should introduce a Hilbert space interpretation with a linear combination of $t_\sE$ and $x_1$ viewed as Euclidean time.   Without essential loss of generality, we can
pick $t_\sE$ as the Euclidean time.   (If instead we choose a direction orthogonal to the $t_\sE-x_1$ plane as the Euclidean time direction, then after Wick rotation to Lorentz signature,
$\P$ will remain as a linear operator, a $\pi$ rotation of two coordinates in Minkowski space.)   

The key fact is now the following.   In a Euclidean correlation function such as a two-point function $\la \phi(t_{\sE},x_{1},\cdots)\phi'(t'_{\sE},x'_{1}\cdots)\ra$, the ordering of
the two operators has no intrinsic meaning.   However, when we interpret such a correlation function in terms of a transfer matrix $e^{-\tau H}$ and a Hilbert space of
physical states, the operators are always ordered in the direction of increasing $t_{\sE}$.   Thus if $t_{\sE}>t'_{\sE}$, the two-point function $\la \phi(t_{\sE},\vec x)\phi'(t'_{\sE},\vec x')\ra$
(where $\vec x=(x_1,\cdots , x_{D-1})$ is the full set of spatial coordinates) is interpreted in operator language as
\be\label{zimbo}\la\Omega|\phi(0,\vec x)e^{-(t_{\sE}-t'_{\sE})H}\phi'(0,\vec x')|\Omega\ra.\ee
Here $\Omega$ is the vacuum state,  $\phi(0,\vec x)$ and $\phi(0,\vec x')$ are Hilbert space operators defined at $t_{\sE}=0$, and in writing the correlation function as in eqn. (\ref{zimbo}), 
we use the fact that by time-translation symmetry, it only depends on the difference $t_\sE-t'_\sE$, not on $t_\sE$ and $t'_\sE$ separately.
   To interpret $\la \phi(t_{\sE},\vec x)\phi'(t'_{\sE},\vec x')\ra$
as the expectation value in the vacuum of a product of Hilbert space states, we have to order the operators as in eqn. (\ref{zimbo}), since the operator $e^{-\tau H}$ is not well-defined for $\tau<0$.   
If $t'_\sE>t_\sE$, then to give a Hilbert space interpretation to the same correlator  $\la \phi(t_{\sE},\vec x)\phi'(t'_{\sE},\vec x')\ra$, we write a similar formula with
the opposite ordering, with $\phi'(0,\vec x')$ on the left.   If $t_\sE=t'_\sE$ (but $\vec x\not=\vec x'$ so that the correlation function we are discussing is well-defined), then the factor
$e^{-(t_\sE-t'_\sE)H}$ drops out and the correlation function has a Hilbert space interpretation with either ordering (telling us that $\phi(0,\vec x)$ and $\phi'(0,\vec x')$ commute
for $\vec x\not=\vec x'$, as expected).   

Now let us give a Hilbert space interpretation to the relation (\ref{zorro}) between correlation functions that follows from $\P$ symmetry.   The key point is that because  $\P$
reverses the sign of $t_\sE$, it reverses the ordering of operators in $t_\sE$.  Therefore, in view of what has just been explained, when we interpret the identity (\ref{zorro})
in terms of operators in Hilbert space, the ordering of the operators is reversed on the right hand side.    Without essential loss of generality, we can assume that the operators are labeled so that $t_{\sE,1}\geq t_{\sE,2}\geq \cdots\geq t_{\sE,d}$.
In that case, the identity between Hilbert space matrix elements that follows from $\P$ symmetry is
\be\label{zorrox} \la \Omega|\phi_1(t_{\sE,1},x_{1,1},\cdots) \cdots \phi_d(t_{\sE,d},x_{1,d},\cdots)|\Omega\ra = \la \Omega|\t\phi_d(-t_{\sE,d},-x_{1,d},\cdots)\cdots \t\phi_1(-t_{\sE,1},-x_{1,1},\cdots)|
\Omega\ra, \ee
with the reverse ordering of the operators on the right hand side.   Eqn. (\ref{zorrox}) can be expressed in terms of operators $\phi_i(0,\vec x_i)$ at $t_\sE=0$  by using time-translation symmetry and inserting judicious
factors of $e^{-\tau X}$, but we omit this as it makes the formulas slightly less transparent.   On both sides of eqn. (\ref{zorrox}), operators are ordered so that $t_\sE$ increases
from right to left.

Now we can ask the following question:   what kind of symmetry operator $\CRT$ must exist in the Hilbert space formulation of the theory  to account for the fact
that the correlation functions satisfy the identity of eqn. (\ref{zorrox})?    In general, we assume that $\CRT$ conjugates fields in a way analogous to the action of $\P$:
\be\label{telfog}\CRT \phi(t_\sE,x_1,x_2,\cdots,x_{D-1})\CRT^{-1} = \h\phi(-t_\sE,-x_1,x_2,\cdots,x_{D-1}). \ee
Here $\phi\to \h\phi$ represents some action of $\CRT$ on the spin labels  that $\phi$ may carry.   The formula (\ref{telfog}) is analogous to eqn. (\ref{undu}) for the action of $\P$,
but the interpretation is different.  In eqn. (\ref{telfog}), $\CRT$ is supposed to be a Hilbert space operator that is a symmetry of the vacuum state
\be\label{gofog} \CRT|\Omega\ra =|\Omega\ra=\CRT^\dagger|\Omega\ra,\ee
and therefore will have consequences for Euclidean correlation functions.  Instead, eqn.  (\ref{undu}) is a rule that corresponds to an invariance of Euclidean correlation functions,
but it does not represent the transformation of $\phi(t_\sE,x_1,\cdots)$ under conjugation by an operator; indeed, eqn. (\ref{undu}) was stated without having introduced any machinery
involving Hilbert spaces and operators.   

What kind of operator satisfying eqn. (\ref{telfog}) and (\ref{gofog}) will imply the expected identity (\ref{zorrox}) among correlation functions?   If we assume that $\CRT$ is a linear
operator, we will be out of luck.   Using (\ref{gofog}) and (\ref{telfog}) (and noting that $\CRT|\Omega\ra=|\Omega\ra$ implies $|\Omega\ra=\CRT^{-1}|\Omega\ra$), we will get
\begin{align}\label{milfog}& \la \Omega|\phi_1(t_{\sE,1},x_{1,1},\cdots) \cdots \phi_d(t_{\sE,d},x_{1,d},\cdots)|\Omega\ra\cr =&\la\CRT^\dagger \Omega|\phi_1(t_{\sE,1},x_{1,1},\cdots) \cdots \phi_d(t_{\sE,d},x_{1,d},\cdots)|\CRT^{-1}\Omega\ra\cr \overset{?}{=}& \la\Omega|\CRT\phi_1(t_{\sE,1},x_{1,1},\cdots)\CRT^{-1}\cdots \CRT\phi_d(t_{\sE,d},x_{1,d},\cdots)\CRT^{-1}|\Omega\ra
\cr= & \la\Omega|\h\phi_1(-t_{\sE,1},-x_{1,1} ,\cdots) \cdots \h\phi_d(-t_{\sE,d},-x_{ 1,d},\cdots)|\Omega\ra. \end{align}
This is almost the identity that we want, assuming that $\h\phi$ and $\t\phi$ coincide, except that in the desired identity (\ref{zorrox}), the order of multiplication of operators
is  reversed on the right hand side.

How can we correct for this?  In eqn. (\ref{milfog}), the step where it was assumed that  $\CRT $ is a linear operator is  indicated by the question mark in $\overset{?}{=}$.
If $\CRT$ is linear, we have $\la\CRT^\dagger\Omega|\chi\ra=\la\Omega|\CRT|\chi\ra$, for any state $\chi$, as has been assumed in eqn. (\ref{milfog}).   Instead, if $\CRT$ is
antilinear, the correct version of the formula is $\la\CRT^\dagger\Omega|\chi\ra=\overline{\la\Omega|\CRT\chi\ra}=\la\CRT\chi|\Omega\ra.$   The analog of eqn. (\ref{milfog}) if
$\CRT$ is assumed to be antilinear is thus a very similar relation, except that in the final result the operators act on $\Omega$ in the bra rather than in the ket:
\begin{align}\label{nilfog}& \la \Omega|\phi_1(t_{\sE,1},x_{1,1},\cdots) \cdots \phi_d(t_{\sE,d},x_{1,d},\cdots)|\Omega\ra\cr = & \la\h\phi_1(-t_{\sE,1},-x_{1,1} \cdots) \cdots \h\phi_d(-t_{\sE,d},-x_{ 1,d}\cdots)
\Omega|\Omega\ra. \end{align}
Of course, operators acting on the bra can be replaced by adjoint operators acting on the ket.   This reverses the order in which the operators are multiplied.   We get
\begin{align}\label{bilfog}& \la \Omega|\phi_1(t_{\sE,1},x_{1,1},\cdots) \cdots \phi_d(t_{\sE,d},x_{1,d},\cdots)|\Omega\ra\cr = & \la\Omega|\h\phi^\dagger_d(-t_{\sE,d},-x_{1,d}, \cdots) \cdots \h\phi^\dagger_1(-t_{\sE,1},-x_{ 1,1},\cdots)|\Omega\ra. \end{align}
We see that this agrees with the expected identity (\ref{zorrox}) if and only if
\be\label{nunzio} \h\phi=\t \phi^\dagger. \ee
Since the operation $\phi\to \t\phi$ is linear, and the operation $\phi\to \phi^\dagger$ is antilinear, eqn. (\ref{nunzio}) says that $\phi\to\h\phi$ is antilinear.
That is as expected, since in the derivation we have assumed that $\CRT$ is antilinear.

The conclusion is that the identity (\ref{zorrox}) that expresses invariance of Euclidean correlation functions under a $\pi$ rotation can be interpreted in a Hilbert space formulation
of the theory in terms of an antilinear operator $\Theta$ whose action on local fields is uniquely determined by eqn. (\ref{nunzio}).     We reached this conclusion without having to
explicitly discuss the analytic continuation to Lorentz signature.   However, at this point we can readily explain the basic idea of that continuation.   First, analytically
continue from real $t_\sE$ to a complex variable $t_\sE=\tau+\i t$, where $\tau$ and $t$ are real.    A two-point function $\la\phi(\tau+\i t,\vec x)\phi'(\tau'+\i t',\vec x')\ra$ 
then can be written in operator language precisely as in eqn. (\ref{zimbo}):
\be\label{pimbo}\la\Omega| \phi(0,\vec x)e^{-(\tau-\tau')H-\i (t-t'))H}\phi'(0,\vec x')|\Omega\ra.\ee
Here  $e^{-(\tau+\i t) H}$ is a well-defined  Hilbert space operator, holomorphic in $\tau+\i t$,  if $\tau$ is non-negative, irrespective of the sign of $t$.  To get Lorentz signature correlation functions,
we take $\epsilon=\tau-\tau'$ to be an infinitesimal positive parameter, so the correlation function is 
\be\label{pimbox}\la\Omega|\phi(0,\vec x)e^{-\epsilon H-\i (t-t'))H}\phi'(0,\vec x')|\Omega\ra.\ee
Here we must take $\epsilon> 0$, but either sign of $t-t'$ is allowed, so although in Euclidean signature operators are ordered from right to left in the order of increasing Euclidean
time, in Lorentz signature any time ordering is possible.   The infinitesimal positive parameter $\epsilon$ that appears in eqn. (\ref{pimbox}) is the same infinitesimal positive parameter
that appears in the Feynman $\i\epsilon$.   For $\epsilon>0$, correlation functions such as the two-point function (\ref{pimbox}) are well-defined  functions, holomorphic in
time differences such as
$\epsilon+\i(t-t')$ between successive operators.  
In the limit $\epsilon\to 0$, the correlation functions may develop singularities and must be interpreted as distributions.     Concerning this statement, we
note that the  most commonly considered local operators in quantum field  theory have singularities at short distances that are bounded by power laws.\footnote{A typical exception is an operator such
as $e^{\phi(t,\vec x)}$ in dimension $D>2$, where $\phi$ is a free scalar field.}  In general, a function $f(\epsilon+\i t)$ that is holomorphic for $\epsilon>0$ and is bounded by a lower law 
for $\epsilon\to 0$ may have singularities at $\epsilon=0$, but its $\epsilon\to 0$ limit always exists as a distribution.\footnote{For an elementary proof, see Proposition 4.2 in
\cite{Fewster}. The converse is also true:  if
$f(\epsilon+\i t)$ is not bounded by a power of $\epsilon$ as $\epsilon\to 0^+$, then it does not have a limit as $\epsilon\to 0$ as a distribution.
  Therefore, for $D>2$, the object $e^{\phi(t,\vec x)}$, whose correlation functions have singularities that are worse than power laws, cannot be interpreted in Lorentz
  signature as an operator-valued distribution.}     Therefore, Lorentz signature correlation functions are well-defined as distributions with any ordering of the operators.

In this discussion, we have implicitly assumed that the operators $\phi_1,\cdots, \phi_d$ are bosonic.   If instead $n=2k$ of them are fermionic, there are a few changes that
actually were essentially already described in section \ref{fermsym}.   Reversing the order of $n$ fermion operators gives a factor $(-1)^{n(n-1)/2}=(-1)^{n/2}=(-\i)^n$.   This is compensated
by including  a factor of $\i$ in the relation (\ref{nunzio}) between the action of $\P$ and of $\CRT$ if the field $\phi$ is fermionic.
\vskip1cm
 \noindent {\it {Acknowledgements}}  
  Research supported in part by NSF Grant PHY-2207584.  I thank D. Freed and D. Harlow for discussions.
 \bibliographystyle{unsrt}

\end{document}